\documentclass[showpacs,showkeys,amsmath,amssymb,12pt]{revtex4}

\usepackage{graphicx}

\begin{document}

\title{Gravitational Lensing by Rotating Naked Singularities}

\author{Galin N. Gyulchev\footnote{E-mail: gyulchev@phys.uni-sofia.bg}, Stoytcho S. Yazadjiev\footnote{E-mail: yazad@phys.uni-sofia.bg} }

\affiliation{Department of Theoretical Physics, Faculty of Physics, Sofia University,\\
5 James Bourchier Boulevard, 1164 Sofia, Bulgaria
}

\date{\today}

\begin{abstract}
We model massive compact objects in galactic nuclei as stationary,
axially-symmetric naked singularities in the
Einstein-massless scalar field theory and study the resulting gravitational lensing. In the weak deflection limit we study analytically the position of the two weak field images, the corresponding signed and absolute magnifications as well as the centroid up to post-Newtonian order. We show that there are a static post-Newtonian corrections to the signed magnification and their sum as well as to the critical curves, which are function of the scalar charge. The shift of the critical curves as a function of the lens angular momentum is found, and it is shown that they decrease slightingly for the weakly naked and vastly for the strongly naked singularities with the increase of the scalar charge. The point-like caustics drift away from the optical axis and do not depend on the scalar charge. In the strong deflection limit approximation we compute numerically the position of the relativistic images and their separability for weakly naked singularities. All of the lensing quantities are compared to particular cases as Schwarzschild and Kerr black holes as well as Janis--Newman--Winicour naked singularities.
\end{abstract}

\pacs{95.30.Sf, 04.20.Dw, 04.70.Bw, 98.62.Sb}

\keywords{Relativity and gravitation; Classical black holes; Naked singularities; Gravitational lensing}

\maketitle

\section{Introduction}

Is there a naked singularity exist or not, it is probably one of the most important unsolved problems in general relativity. According to the standpoint of the cosmic censorship conjecture the spacetime singularities of gravitational collapse are hidden within black holes \cite{RP1, RMWald} and therefore can not be observed. In this work, we model the massive, dark, radio object Sgr A$^\ast$ in the center of the Galaxy as a rotating generalization of the Janis--Newman--Winicour naked singularity. Using the predictions of the gravitational lensing caused by the space-time under consideration we obtain the physical parameters of the rotating gravitational lens, which uncover the rotating naked singularities concealed in the black holes.

In the last decades the gravitational lensing attracted ever more and more the interests of the science community. As result the theory have developed in two frames. The former examine the problem for the distribution of a photon on a radial distance much larger than the gravitational radius of the lens. In this case the light ray has a small deflection angle, therefore we are speaking about the gravitational lensing in the \textit{weak deflection limit}. Then two weak field images on the each side of the lens appeared. The letter discuss the photons winding many times closely around the lens before reaching into the observer. This process is also known as a gravitational lensing in the \textit{strong deflection limit}. I this case, an infinite series of highly demagnified relativistic images on both sides of the lens shadow should be appeared.

It is known that the lensing observables might be successfully described by a perturbation theory applied to general relativity. The theory of gravitational lensing in the weak deflection limit approximation has been developed for a Schwarzschild point-mass lens \cite{SEF, PettersLevine}. The rotational case has been studied for the first time up to a post-Newtonian expansion by Epstein and Shapiro \cite{EpsteinShapiro} and after that by Richter and Matzner \cite{RichterMatzner}. Later on, Bray \cite{Bray} investigated the multi-imaging aspect of Kerr black hole lensing as resolved the equations of motion for a light ray up to and including second order terms in scaled black hole mass $m/r_{min}$ and angular momentum $a/r_{min}$, where $r_{min}$ is the distance of closest approach. Gravitational lensing by rotating stars has been considered by Glinstein \cite{Glinstein} and later on by Sereno \cite{Sereno1}. The weak field Reissner-Nordstrom black hole lensing has been done by Sereno \cite{Sereno2}. The gravitatonal-magnetic effect in the propagation of light in the field of self-gravitating bodies has been investigated by Kopeikin \textit{at al.} \cite{Kopeikin}. Asada and Kasai \cite{AsadaKasai} have found that up to the first order in the gravitational constant $G$, a rotating lens is not distinguishable from a not-rotating one. They have found that because of the global translation of the center of lens mass the Kerr lens is observationally equivalent to the Schwarzschild one at linear order in mass $m$ and the specific lens angular momentum $a$. Later on, Asada, Kasai, and Yamamoto have shown \cite{AsadaKasaiYamamoto}, that the nonlinear coupling break the degeneracy so that the rotational effect becomes in principle separable for multiple images or a single source. After that, Sereno \cite{Sereno3} considered a gravitational lensing in metric theories of gravity in post-post-Newtonian order with gravitomagnetic field. Recently, Keeton and Petters \cite{KeetonPetters} have developed a general formalism for lensing by spherically symmetric lenses up to post-post-Newtonian order. Sereno and de Luca \cite{SerenoLuca} extended their approach to the case of the Kerr black hole. Finally, Werner and Petters \cite{WernerPetters} based on the analysis by Asada, Kasai, and Yamamoto, applied a simpler method to derive image positions and magnifications up to post-Newtonian order. Using the degeneracy in the case of the Kerr black hole lensing in the weak deflection limit, they presented lensing observables for the two weak field images in post-Newtonian terms with scaled lens angular momentum $a/m$.

In 1959 Darwin \cite{Dar} examined a photon trajectories passing in the vicinity of the black hole and showed the larger deflection, which suffer a light ray without falling into the event horizon. This effect was considered again also in \cite{LON}. After that Fritteli, Kling and Newman \cite{FKN} as well as Virbhadra and Ellis \cite{Vir1} developed this idea and worked out a definition of an exact lens equation. Virbhadra, Nirasimha and Chitre \cite{Vir3} as well as  Virbhadra and Ellis \cite{Vir2} studied later numerically the lensing by static and spherically symmetric naked singularity and showed the influence of the scalar field to the lensing observables. In contrast to this Perlick \cite{Per} has considered lensing in a spherically symmetric and static spacetime, based on the lightlike geodesic equation without approximations. Bozza \cite{Bozza1} developed an analytical technique based on the larger deflection of the light ray and showed that the deflection angle diverges logarithmically as light rays approach the photon sphere of a Schwarzschild black hole. The expectation that the relativistic images should be test the gravity in the strong deflection limit has leaded to application of this method to various metrics from the general relativity, string and alternative theories \cite{Group1, Group2, Group3, Eiroa, MukherjeeMajumdar}. Analytically was studied and the Kerr black hole lensing in Refs. \cite{Group5} and extended to Kerr-Sen black hole lensing \cite{GyulchevYazadjiev}. The time delay calculation for higher order images have done by Bozza and Mancini \cite{BozzaMancini1} and Bozza and Sereno \cite{BozzaSereno}. Recently, Vibhadra and Keeton \cite{VirbhadraKeeton} have examined numerically the time delay and magnification centroid due to gravitational lensisng by black holes and naked singularities. Out of the assumption for the very far source positions Bozza and Scarpetta \cite{BozzaScarpetta} have considered black hole gravitational lensing with arbitrary source distances with respect to the black hole.

In our desire to bridge the gap between the weak and the strong deflection limit analysis of gravitational lensing we can refer to Amore and Arceo \cite{AmoreArceo}. They present a method which can be used to obtain arbitrarily, accurate, analytical expressions for the deflection angle of light propagating in a given metric. An effective analytical formalism for the Schwarzschild deflection angle which describe with satisfactorily accuracy both weak and strong deflection series have also been developed by Iyer and Petters \cite{IyerPetters}.

The purpose of this paper is to consider the gravitational lensing
by rotating naked singularities and to explore how it
differs from the Kerr black hole lensing. The weak gravitational lensing allow us to describe the light ray trajectory, where the closest approach distance $r_{0}$ and the impact parameter $J$ both lie outside the gravitational radius $r_{g}=2M$. Besides, the strong gravitational lensing could provide profound
examination of the space--time around different kinds of black holes
and naked singularities. Therefore, following \cite{Bozza1, WernerPetters} in the present work we wish to study gravitational
lensing in the weak and strong deflection limit due to a stationary,
axially-symmetric weakly naked singularities and to compare the
results to rotating strongly naked singularities with the
aim of investigating the influence of the scalar field on
the behavior of the bending angle, on the position of the
images and on their magnification as well as in the
critical curves and caustics.

The outline of this paper is as follows. The second section contains
a description of a Kerr-like solution of the
Einstein-massless scalar field equations. In Sec. III we
discuss the full lens equation. In Sec. IV a gravitational lensing in weak deflection limit by Kerr black hole, weakly and strongly naked singularity is investigated and the critical curves and the caustic structure are considered. In Sec. V the deflection angle is numerically computed in the equatorial plane and its dependence from the scalar charge and the lens angular momentum is shown. In Sec. VI we discuss the equatorial lensing by Kerr black hole, weakly naked singularity and marginally strongly naked singularity in the strong deflection limit and compute the positions of the relativistic images and their separability. A discussion of the results is given in Sec. VII.

\section{Rotating singularity spacetime}

We consider a Kerr-like solution \cite{KB} to the
Einstein-massless scalar equations
($R_{ij}=8\varphi_{,i}\varphi_{,j}$ with $\varphi_{,i}^{;i}=0$,
where $R_{ij}$ is the Ricci tensor and $\varphi$ is the
massless scalar field). This solution is rotating generalization of
the Janis--Newman--Winicour (see for example \cite{VirJNW}) solution  and is given by the line
element
\begin{eqnarray}\label{KLM}
   ds^{2}&=&\left(1-\frac{2Mr}{\gamma\rho}\right)^{\gamma}(dt-w d\phi)^{2} \nonumber \\
  &-&\left(1-\frac{2Mr}{\gamma\rho}\right)^{1-\gamma}\rho\left(\frac{dr^{2}}{\Delta}+d\vartheta^{2}+\sin^{2}{\vartheta}d\phi^{2}\right)   \nonumber \\ &+&2w(dt-w d\phi)d\phi, \end{eqnarray}
and the scalar field
\begin{eqnarray}
    \varphi=\frac{\sqrt{1-\gamma^{2}}}{4}\ln{\left(1-\frac{2Mr}{\gamma\rho}\right)},
\end{eqnarray}
where
\begin{eqnarray}
     \gamma=\frac{M}{\sqrt{M^{2}+q^2}}, \,\,\,\,\, w=a\sin^{2}\vartheta, \,\,\,\,\, \rho=r^{2}+a^{2}\cos^{2}\vartheta, \,\,\,\,\, \Delta=r^{2}+a^{2}-\frac{2Mr}{\gamma}.
\end{eqnarray}
$M$ and $q$, the Arnowitt-Deser-Misner (ADM) mass and scalar charge, are constant real parameters in this solution. $a=L/M$
is the angular momentum of the rotating object in units of the mass. Hereafter we will not consider the massless scalar field case (\textit{e. i.} $\gamma=0$ or $q/M=\infty$). For $a=0$ and $q\neq0$ this solution reduces to
the Janis--Newman--Winicour solution, for $a\neq0$ and $q=0$ to the Kerr black hole, while in the particular
case $a=0$ and $q=0$ it reconstructs the Schwarzschild solution. The above solution is in fact the Einstein-frame version
of the original Jordan-frame solution to the Brans-Dicke equations found in \cite{KB}.

In the case $\gamma=1$ ($q=0$) when  the Kerr black hole solution is recovered, there is an event horizon with spherical topology,
which is the biggest root of the equation $\Delta=0$ and is given by
\begin{equation}
  r_{H}=M+\sqrt{M^{2}-a^{2}}
\end{equation}
for $|a|\leq{M}$. Beyond this critical value of the spin there is no
event horizon and causality violations are present in the whole
space-time, with the appearance of a naked singularity. The
ergosphere is defined as a surface on which the Killing vector
$\frac{\partial}{\partial{t}}$ is isotropic, \textit{i.e}
$g_{tt}=0$. So the ergosphere lies at
\begin{equation}
  r_{es}=M+\sqrt{M^{2}-{a}^{2}\cos^{2}\vartheta}.
\end{equation}
For $0<\gamma<1$ the metric (\ref{KLM}) describes  rotating naked singularities with mass $M$ and angular momentum $Ma$. In order to see
that the solution is singular for  $0<\gamma<1$ we  calculate the Ricci scalar curvature and find
\begin{eqnarray}
 R=\frac{2(\gamma^{2}-1)M^{2}}{\gamma^{2}\rho^{5}}\left(1-\frac{2Mr}{\gamma\rho}\right)^{\gamma-3}[\Delta(r^{2}-a^{2}\cos^{2}\vartheta)^{2}+(ra^{2}\sin^{2}2\vartheta)^{2}].
\end{eqnarray}
As it is seen, the scalar curvature diverges where $g_{tt}=0$ which shows the presence of a curvature singularity at
\begin{equation}
  r_{cs}=\frac{1}{\gamma}\left[M+\sqrt{M^{2}-\gamma^{2}{a}^{2}\cos^{2}{\vartheta}}\right],
\end{equation}
as the domain of variation of $\theta$ depends on the ratio $\gamma^2a^2/M^2$. One can show that there is at least one equatorial null geodesic with one end on the singularity and the other on the future null infinity i.e.
the singularity is indeed naked. The global naked singularity nature of the static JNW solution for $0\leq\gamma<1$ was first shown in \cite{VJJ} (see also the Seifert conjecture for naked singularities \cite{VirS}).

\section{Equation of the gravitational lens}

The equation of the gravitational lens, which allows
small as well as large bending of the light ray is \cite{Vir1}
\begin{equation}\label{LensEq1}
    \tan\boldsymbol{\mathcal{B}}=\tan\boldsymbol{\Theta}-\frac{D_{LS}}{D_{OS}}[\tan\boldsymbol{\Theta}+\tan(\boldsymbol{\tilde{\alpha}}-\boldsymbol{\Theta})],
\end{equation}
where $D_{LS}$ and $D_{OS}$ respectively are the
lens-source and the observer-source angular diameter distances.
The optical axis of the gravitational lens systems connects the origins of Cartesian angular coordinates $\boldsymbol{\mathcal{B}}=(\mathcal{B}_{1}, \mathcal{B}_{2})$, $|\boldsymbol{\mathcal{B}}|=\mathcal{B}$ and $\boldsymbol{\Theta}=(\Theta_{1}, \Theta_{2})$, $|\boldsymbol{\Theta}|=\Theta$ respectively of the source plane and the lens plane.
Given a source position $\boldsymbol{\mathcal{B}}$, the values of $\boldsymbol{\Theta}$,
that solve this equation, give the pos  ition of the observed
images measured from the optical axis. The projections of the deflection angles into the lens plane are denoted by $\boldsymbol{\tilde{\alpha}}=(\tilde{\alpha_{1}}, \tilde{\alpha_{2}})$, $|\boldsymbol{\tilde{\alpha}}|=\tilde{\alpha}$. For the impact
parameter of the light ray we have $J=D_{OL}\sin\Theta$, where $D_{OL}$ is the observer-lens angular diameter distance. For small angles, Eq. (\ref{LensEq1}) reduces to the well known in the literature weak field lens equation, as well as to the strong deflection limit lens equation \cite{BozQuazi}. For spherically symmetric lenses we refer to \cite{BS}, where most general lens equation is constructed.

As we shall see below, we will solve the lens equation in the weak deflection limit in the cases of presence and absence of the photon sphere in order to explore the differences in the
gravitational lensing by various of naked singularities. Moreover, since the strong deflection limit
allows a simple analytical investigation of the gravitational
lensing properties, after calculating of $\tilde{\alpha}$
for the rotating singularity space-time in this approximation, we will solve the
lens equation in order to derive the relativistic images when a
photon sphere exists \cite{BozQuazi}.

\section{Gravitational lensing by Rotating Naked Singularities in the weak deflection limit.}

\subsection{Image positions}

In this part we will study the gravitational lensing by rotating naked singularities in the weak field regime up to the post-Newtonian order. We will distribute the source and the observer in the asymptotically flat region in such a way that the source is situated beyond the lens plane. Following the scheme exposed in \cite{WerPet} up to the post-Newtonian limit, we can express the angular coordinates of the source and the image in Einstein ring scale in the following way
\begin{eqnarray}
  \mathcal{B}&=&\theta_{E}\beta=\theta_{E}\left(\beta_{(0)}+\beta_{(1)}\epsilon+\mathcal{O}(\epsilon^2)\right), \\
  \Theta&=&\theta_{E}\theta=\theta_{E}\left(\theta_{(0)}+\theta_{(1)}\epsilon+\mathcal{O}(\epsilon^2)\right),
\end{eqnarray}
where angular radius of the Einstein ring and the expansion parameter are
\begin{equation}
    \theta_E=\sqrt{\frac{4MD_{LS}}{D_{OL}D_{OS}}},\,\,\,\,\,\,\,\,\ \epsilon=\frac{\theta_{E}D_{OS}}{4D_{LS}}.
\end{equation}
The Janis--Newman--Winicour deflection angle is
\begin{equation}
    \tilde{\alpha}=\frac{4M}{J}+\frac{4M^2}{J^2}\left[ 1-\frac{1}{16\gamma^2} \right]\pi+\mathcal{O}\left(\frac{M^3}{J^3}\right).
\end{equation}
We refer to \cite{Vir3, VirbhadraKeeton} for different representations of the deflection angle.

Then up to the post-Newtonian order the weak field Janis--Newman--Winicour lens equation gets the form
\begin{equation}\label{WFLE}
    \beta=\theta-\frac{1}{\theta}-\left[1-\frac{1}{16\gamma^2}\right]\pi\frac{\epsilon}{\theta^2}.
\end{equation}

Hence, we can advance to the image positions. Let us orient lens coordinates such that the axis $\theta_{2}$ is along the projected lens angular momentum and the axis $\theta_{1}$ is perpendicular to the optical axis. Following \cite{AsadaKasai, AsadaKasaiYamamoto, SerenoLuca}, one can show that up to the post-Newtonian order lensing by rotating naked singularities under consideration is equivalent to lensing by Janis--Newman--Winicour lens but shifted by
\begin{equation}\label{Shift}
    \delta\boldsymbol{\theta}=\theta_{E}(\delta\theta_{1},0)=\theta_{E}(\delta\theta_{1(1)}\epsilon,0),\,\,\,\,\,\,\,\,\ \delta\theta_{1(1)}=\frac{a\sin\vartheta_{O}}{M}.
\end{equation}
Then we can use the lens equation (\ref{WFLE}) to describe the image properties \cite{WerPet}. Because of this shift (\ref{Shift}) the image position $\theta_{1}$ of a direct photon ($a>0$) translates to position $\theta_{1}-\delta\theta_{1}$. By analogy the retrograde image position ($a<0$) is $\theta_{1}+\delta\theta_{1}$. Based on this, allowing positive and negative angular momenta, via the substitution $\theta_{1}\mapsto\theta_{1}-\delta\theta_{1}$ for the two kinds of photons, one can compute the deflection angle $\tilde{\alpha}=\tilde{\alpha}(\theta)$ again and rewrite the lens equation. Thereby, the rotating singularity lens equations are
\begin{eqnarray}\label{WFLE2}
  \beta_{1}&=&\theta_1-\frac{\theta_1-\delta\theta_1}{(\theta_1-\delta\theta_1)^2+\theta_{2}^{2}}-\left[ 1-\frac{1}{16\gamma^2} \right]\pi\frac{\theta_1-\delta\theta_1}{((\theta_1-\delta\theta_1)^{2}+\theta_{2}^{2})^{3/2}}\epsilon+\mathcal{O}(\epsilon^2), \nonumber \\
  \beta_{2}&=&\theta_2-\frac{\theta_2}{(\theta_1-\delta\theta_1)^2+\theta_{2}^{2}}-\left[ 1-\frac{1}{16\gamma^2} \right]\pi\frac{\theta_2}{((\theta_1-\delta\theta_1)^{2}+\theta_{2}^{2})^{3/2}}\epsilon+\mathcal{O}(\epsilon^2).
\end{eqnarray}

At Newtonian order ($\epsilon=0$, $\delta\theta_{1}=\mathcal{O}(\epsilon)$) Eq. (\ref{WFLE2}) reduces to the Janis--Newman--Winicour lens equation (\ref{WFLE}).
The solution can be expanded in a series by $\epsilon$,
\begin{eqnarray}
  \theta_1 &=& \theta_{1(0)}+\theta_{1(1)}\epsilon+\mathcal{O}(\epsilon^2), \nonumber \\
  \theta_2 &=& \theta_{2(0)}+\theta_{2(1)}\epsilon+\mathcal{O}(\epsilon^2).
\end{eqnarray}
The square of the solution of Newtonian order lens equation is $\theta_{(0)}^{2}=\theta_{1(0)}^{2}+\theta_{2(0)}^{2}$, where
\begin{eqnarray}\label{Theta_Sch}
  \theta_{1(0)}^{\pm} &=& \frac{\beta_{1}}{2}\left(1\pm\sqrt{1+\frac{4}{\beta^2}}\right), \nonumber \\
  \theta_{2(0)}^{\pm} &=& \frac{\beta_{2}}{2}\left(1\pm\sqrt{1+\frac{4}{\beta^2}}\right).
\end{eqnarray}
The square of the scaled angular source position is $\beta^{2}=\beta_{1}^2+\beta_{2}^2$. After deriving the post-Newtonian order correction of the lens equations (\ref{WFLE2}) we find the second terms of the image positions
\begin{eqnarray}\label{Theta_11}
  \theta_{1(1)} &=& \left[1-\frac{1}{16\gamma^2}\right]\frac{\pi\theta_{1(0)}}{(1+\theta_{(0)}^2)\theta_{(0)}}+\frac{(1-\theta_{1(0)}^2+\theta_{2(0)}^2)\delta\theta_{1(1)}}{1-\theta_{(0)}^4}, \nonumber \\
  \theta_{2(1)} &=& \left[1-\frac{1}{16\gamma^2}\right]\frac{\pi\theta_{2(0)}}{(1+\theta_{(0)}^2)\theta_{(0)}}-\frac{2\theta_{1(0)}\theta_{2(0)}\delta\theta_{1(1)}}{1-\theta_{(0)}^4}, \end{eqnarray}
which reduce to the correction terms expected for rotating lenses \cite{WerPet} in the case of Kerr black hole ($\gamma=1$). The post-Newtonian corrections of the positive and negative parity images can be found since we already know $\theta_{1(0)}^{\pm}$, $\theta_{2(0)}^{\pm}$.

\subsection{Critical curves and caustics.}

The critical curves separate the regions in the lens plane where the Jacobian determinant $\tilde{J}$ of the lens map  has opposite sign. For a point lens at these curves, the magnification factor of the images $\mu$ diverges. Formally looking, the critical curves are solution to the equation
\begin{equation}\label{CritCurves}
       \tilde{J}=\frac{\partial\beta_{1}}{\partial\theta_{1}}\frac{\partial\beta_{2}}{\partial\theta_{2}}-
        \frac{\partial\beta_{1}}{\partial\theta_{2}}\frac{\partial\beta_{2}}{\partial\theta_{1}}=0.
\end{equation}
According to the lens Eqs. (\ref{WFLE2}) and (\ref{CritCurves}) up to the post-Newtonian order the Jacobian is
\begin{equation}\label{Jacobian}
    \tilde{J}=1-\frac{1}{\theta_{(0)}^4}+\left(\left[1-\frac{1}{16\gamma^2}\right]\frac{\pi(1
    -\theta_{(0)}^2)^2}{(1+\theta_{(0)}^2)\theta_{(0)}^{5}}-\frac{4\theta_{1(0)}\delta\theta_{1(1)}}{(1+\theta_{(0)}^2)\theta_{(0)}^4}\right)\epsilon+\mathcal{O}(\epsilon^2),
\end{equation}
which reduces to the correction terms expected for Kerr black hole \cite{SerenoLuca} up to post-Newtonian order.

We look for a parametric solution up to the post-Newtonian order in the form
\begin{eqnarray}\label{Caust}
  \theta_{1}^{cr} &=&\frac{\Theta_{1}^{cr}}{\theta_{E}}=\cos\varphi\{1+\delta\theta_{E}(\varphi)\epsilon+\mathcal{O}(\epsilon^2)\},  \nonumber \\
  \theta_{2}^{cr} &=&\frac{\Theta_{2}^{cr}}{\theta_{E}}=\sin\varphi\{1+\delta\theta_{E}(\varphi)\epsilon+\mathcal{O}(\epsilon^2)\},
  \end{eqnarray}
where $\varphi$ is an angle in a polar coordinate system taken in the lens plane with origin at the lens. In that system $\tan\varphi=\tan\Theta_{1}{/}\tan\Theta_{2}$. The first term of (\ref{Caust}) gives the Schwarzschild black hole Einstein ring with radius $\theta_{E}$. Solving Eq. (\ref{CritCurves}) we obtain the deviation coefficient
\begin{equation}\label{DC}
    \delta\theta_{E}=\left[1-\frac{1}{16\gamma^2}\right]\frac{\pi}{2}+\frac{a\sin\vartheta_{O}}{M}\cos\varphi+\mathcal{O}(\epsilon^2).
\end{equation}
For Kerr black hole lensing ($\gamma=1$), (\ref{DC}) reduces to the result of Sereno and Luca \cite{SerenoLuca}.
Then, for an observer set in position $\vartheta_{O}$ and fixed values of $a$ and $\gamma$ a critical curve in the $\{\tan\Theta_{1},\tan\Theta_{2}\}$ plane exists. The positions of the equatorial cross sections of the critical curves are shifted with respect to the static case by
\begin{equation}\label{CritShift}
   \tan\delta\Theta^{cr}\simeq\frac{a\sin\vartheta_{O}}{D_{OL}}+\mathcal{O}(\epsilon^2).
\end{equation}
%----------------------------------------------------------------------------------------
\begin{figure}
   \includegraphics[width=1\textwidth]{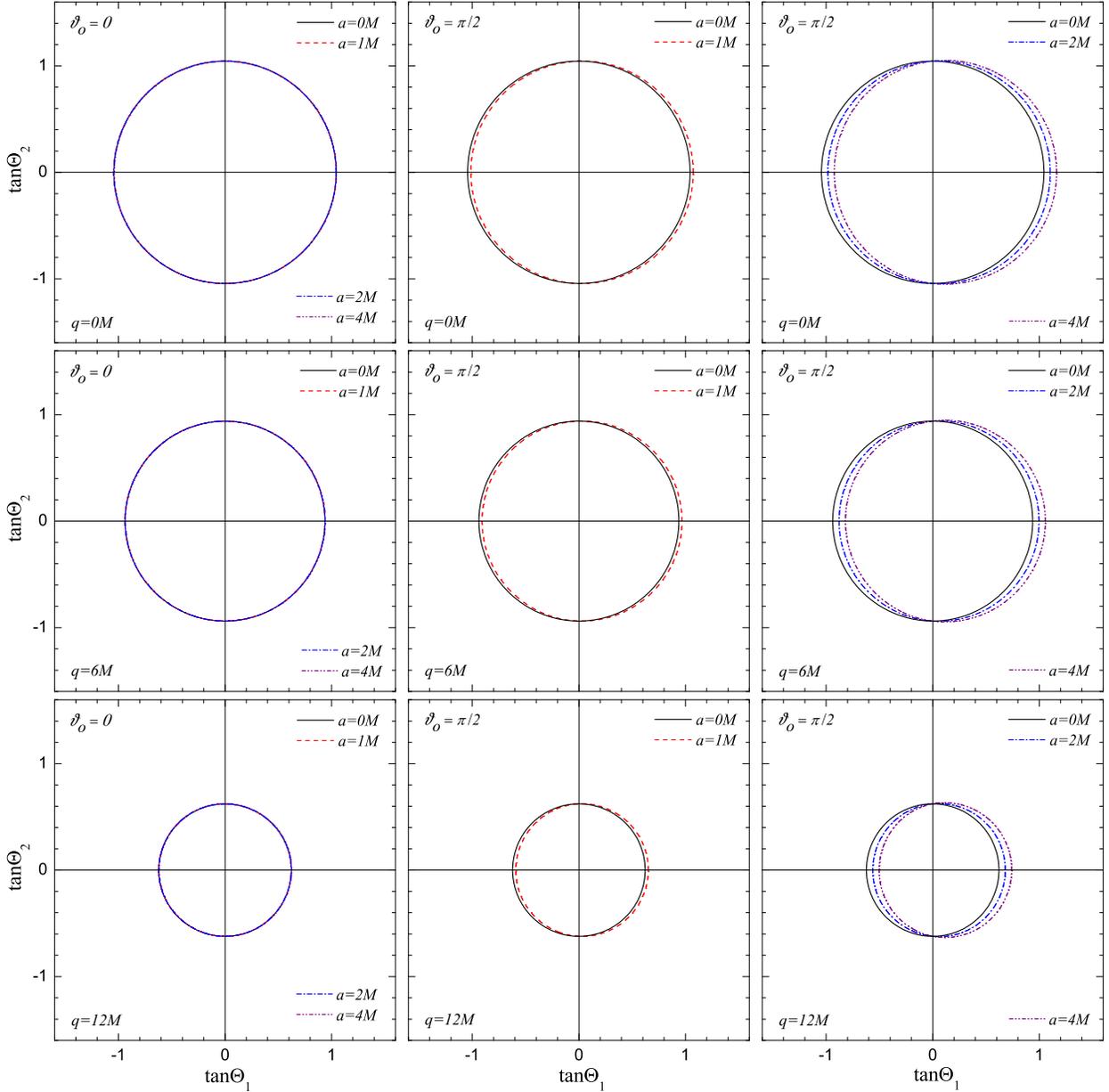}
   \caption{\small Critical curves in the plane $\{\tan\Theta_{1},\tan\Theta_{2}\}$ for a observer-lens position $D_{OL}=7.62$ kpc and lens-source position $D_{LS}=4.85\times10^{-5}$ pc. The Schwarzschild ($a=0M$ and $q=0M$) and the extremal Kerr black hole lens ($a=1M$ and $q=0M$) as well as the Janis--Newman--Winicour ($a=0M$ and $q=6M$, $12M$) and the rotating naked singularity lenses ($a=2M$, $4M$ and $q=6M$, $12M$) are considered. The observer is polar $\vartheta_{O}=0$ and equatorial $\vartheta_{O}=\pi/2$. Axis-lengths are in units of tangent of Einstein angle, $\theta_{E}\simeq157$ $\mu$arcsec.} \label{CCurves}
\end{figure}
%----------------------------------------------------------------------------------------
In order to calculate the caustics we have to find the corresponding source positions. Through lens equations (\ref{WFLE2}) and (\ref{Caust}), up to $\mathcal{O}(\epsilon^3)$ order the caustic are point-like and are positioned in
\begin{equation}\label{Caustic}
    \{ \tan\mathcal{B}_{1}^{cau}, \tan\mathcal{B}_{2}^{cau} \} \simeq \left\{ \frac{a\sin\vartheta_{O}}{D_{OL}}+\mathcal{O}(\epsilon^3), 0 \right\}.
\end{equation}
Our results for the critical curves shift (\ref{CritShift}) and the caustic positions (\ref{Caustic}) coincides with those found in \cite{SerenoLuca} up to the post-Newtonian order.

Critical curves are plotted in Fig. \ref{CCurves} for some values of the scalar charge, the lens angular momentum and the observer's positions. We model the massive dark object in the center of our galaxy as a Kerr black hole and as a rotating generalization of the JNW naked singularities. We assume a point source and set the lens between the source and the observer as assuming that $D_{OL}=7.62$ kpc. Studying the influence of the lens parameters over the critical curves, as an illustration we made all graphics for the source position $D_{LS}=4.85\times10^{-5}$ pc. According to \cite{Eis} the lens has a mass $M=3.61 \times 10^6 M_{\odot}$. In this situation the expansion parameter $\epsilon=0.029846679$.

\subsection{Magnification.}

In the approximation of geometrical optics gravitational lensing causes a change in the cross section of a bundle of light rays, such that the surface brightness is conserved. Therefore, the ratio between the angular area element of the image in the celestial sky, $d\Theta_{1}d\Theta_{2}$, and the angular area element of the source in absence of the lens, $d\mathcal{B}_{1}d\mathcal{B}_{2}$, gives the signed magnification
\begin{equation}\label{SignMagn1}
    \mu=\tilde{J}^{-1}=\left[\frac{\partial\beta_{1}}{\partial\theta_{1}}\frac{\partial\beta_{2}}{\partial\theta_{2}}-
                                  \frac{\partial\beta_{1}}{\partial\theta_{2}}\frac{\partial\beta_{2}}{\partial\theta_{1}} \\
                               \right]^{-1}.
\end{equation}

Disposing with the Eq. (\ref{WFLE2}) the calculations yield
\begin{equation}\label{SignMagn2}
    \mu=\frac{\theta_{(0)}^4}{\theta_{(0)}^4-1}-\left(\left[1-\frac{1}{16\gamma^2}\right]\frac{\pi\theta_{(0)}^3}{(1+\theta_{(0)}^2)^3}-\frac{4\theta_{(0)}^4\theta_{1(0)}\delta\theta_{1(1)}}{(1-\theta_{(0)}^2)^2(1+\theta_{(0)}^2)^3}\right)\epsilon+\mathcal{O}(\epsilon^2),
\end{equation}
which describes the signed magnification for both images. Eq. (\ref{SignMagn2}) reduces to the signed magnification of the Kerr black hole images \cite{WerPet}, when $\gamma=1$.

The individual magnifications of the positive and the negative parity image up to post-Newtonian order can be calculated using (\ref{Theta_Sch}) and (\ref{SignMagn2}). They are respectively
\begin{eqnarray}\label{MuPlasMinus}
  \mu^{+} &=& \frac{(\beta+\sqrt{\beta^2+4})^4}{(\beta+\sqrt{\beta^2+4})^4-16}-\frac{1}{(4+\beta^2)^{3/2}}\left(\left[1-\frac{1}{16\gamma^2}\right]\pi-\frac{4\beta_{1}\delta\theta_{1(1)}}{\beta^3}\right)\epsilon+\mathcal{O}(\epsilon^2), \nonumber \\
  \mu^{-} &=& \frac{(\beta-\sqrt{\beta^2+4})^4}{(\beta-\sqrt{\beta^2+4})^4-16}-\frac{1}{(4+\beta^2)^{3/2}}\left(\left[1-\frac{1}{16\gamma^2}\right]\pi+\frac{4\beta_{1}\delta\theta_{1(1)}}{\beta^3}\right)\epsilon+\mathcal{O}(\epsilon^2). \end{eqnarray}
Than the sum of the signed magnification for the rotating singularity lens, which has the form
\begin{equation}\label{MagnSum}
    \mu_{+}+\mu_{-}=1-\frac{2\pi}{(4+\beta^2)^{3/2}} \left[1-\frac{1}{16\gamma^2}\right]\epsilon+\mathcal{O}(\epsilon^2),
\end{equation}
does not depend on the specific angular momentum $a$ and is equivalent to the result for Janis--Newman--Winicour lens to post-Newtonian order. At order $\mathcal{O}(\epsilon)$ the deviations from the magnification invariant for the Janis--Newman--Winicour lens and the lens under consideration are the same.
%----------------------------------------------------------------------------------------
\begin{figure}
    \includegraphics[width=0.9\textwidth]{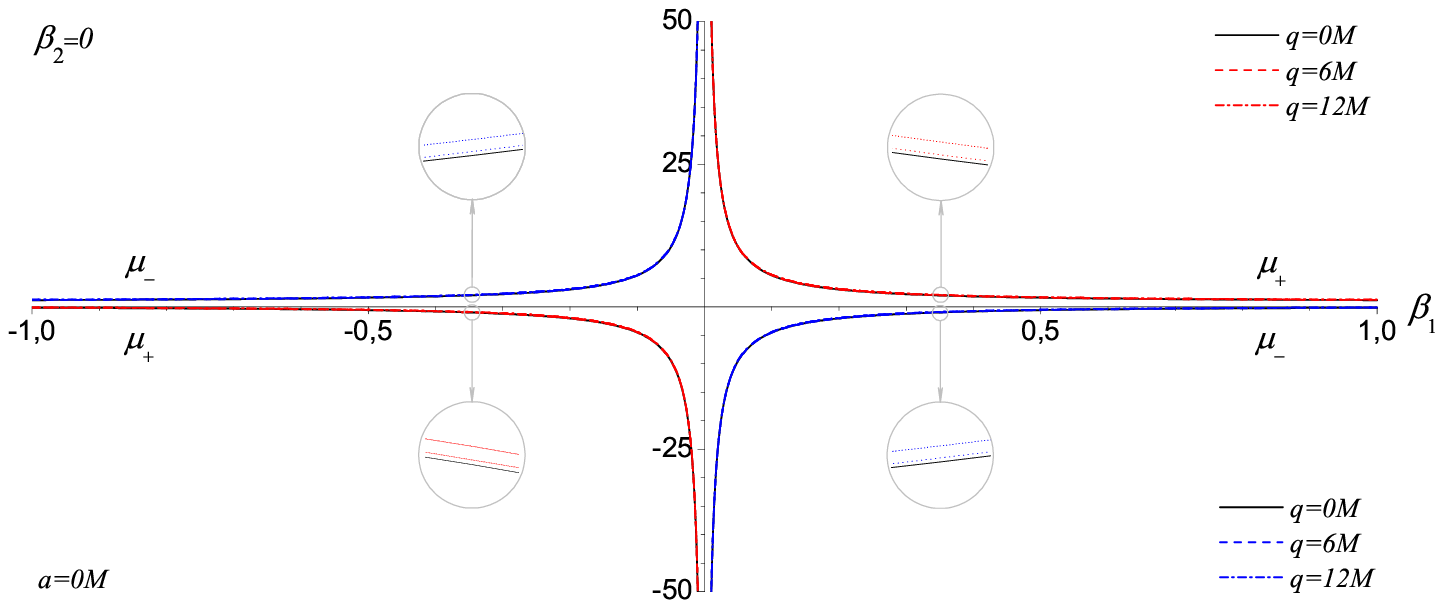} \\
    \vspace{0.45cm}
    \includegraphics[width=0.9\textwidth]{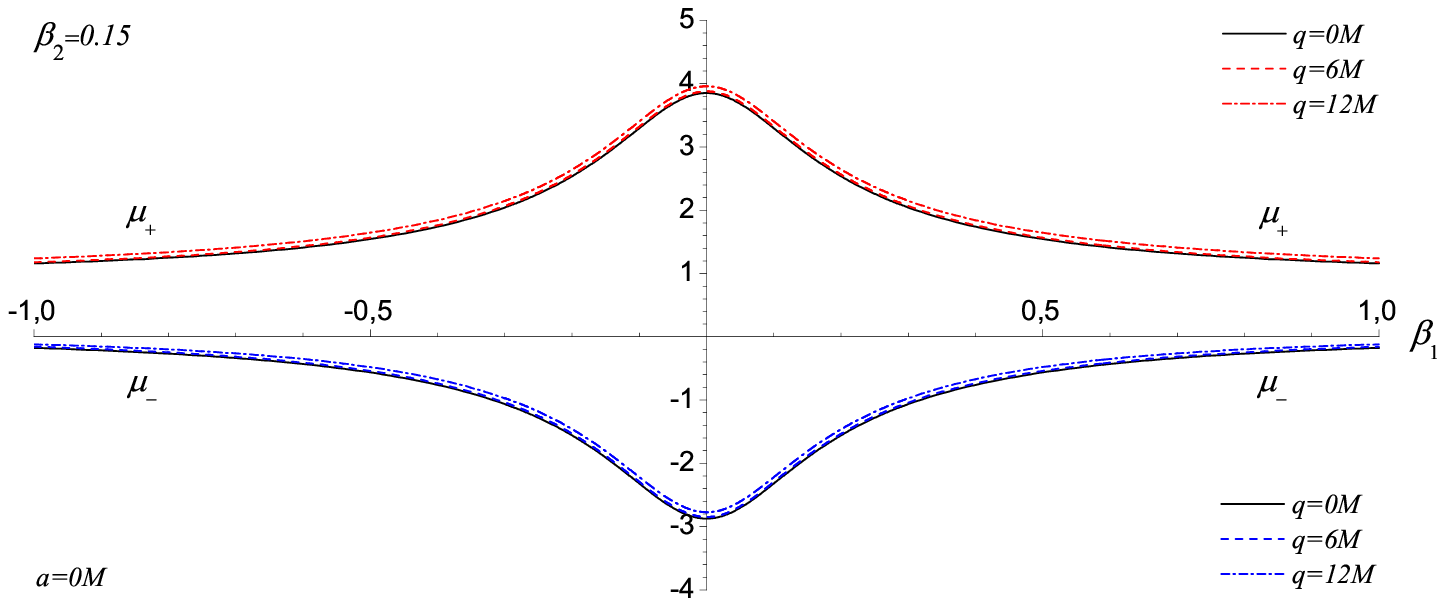} \\
    \vspace{0.45cm}
    \includegraphics[width=0.9\textwidth]{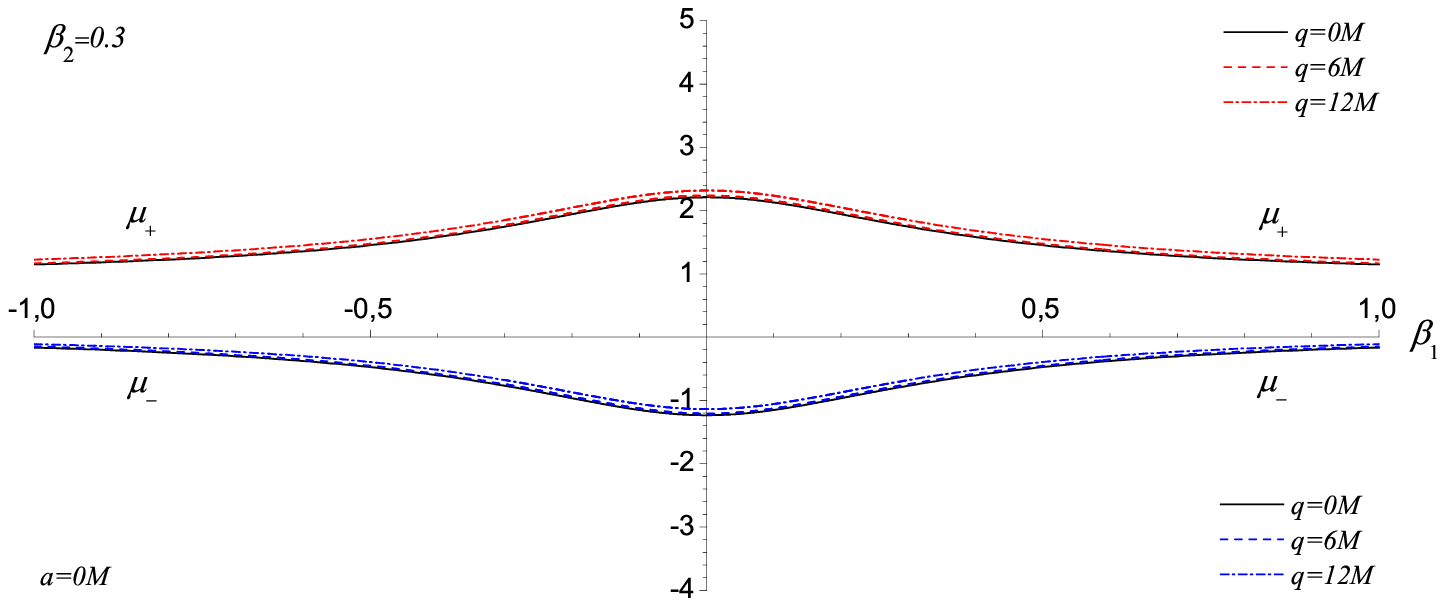}
    \caption{\small The magnification of the positive parity image $\mu_{+}$ and the value of that of the negative parity image $\mu_{-}$ as a function of the scaled angular coordinate $\beta_{1}$ of the source for different scaled angular coordinate $\beta_{2}$. The Schwarzschild black hole (solid line) and the Janis--Newman--Winicour  naked singularity lenses (dashed line; dashed doted line) are considered. (a) The observer is equatorial $\vartheta_{O}=\pi/2$ at position $D_{OL}=7.62$ kpc and the source is at position $D_{LS}=4.85\times10^{-5}$ pc. (b) The abscissa is in units of Einstein angle, $\theta_{E}\simeq157$ $\mu$arcsec.} \label{SM1}
\end{figure}
%----------------------------------------------------------------------------------------
%----------------------------------------------------------------------------------------
\begin{figure}
    \includegraphics[width=0.9\textwidth]{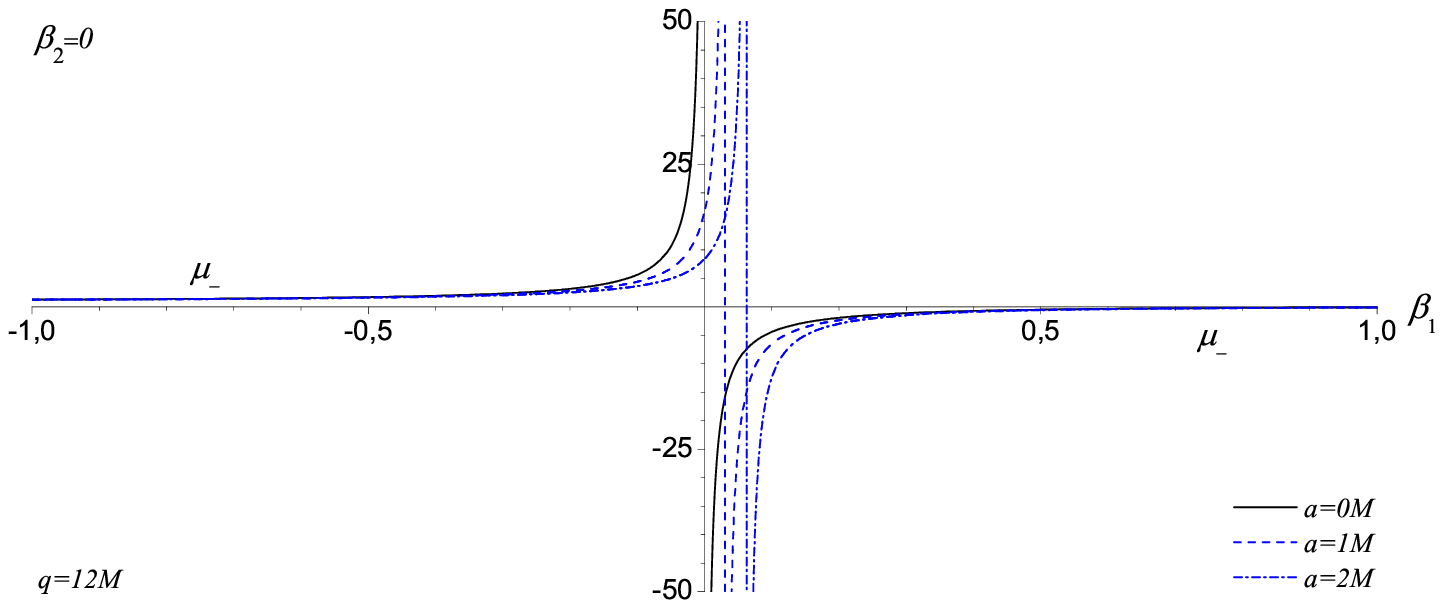} \\
    \vspace{1.0cm}
    \includegraphics[width=0.9\textwidth]{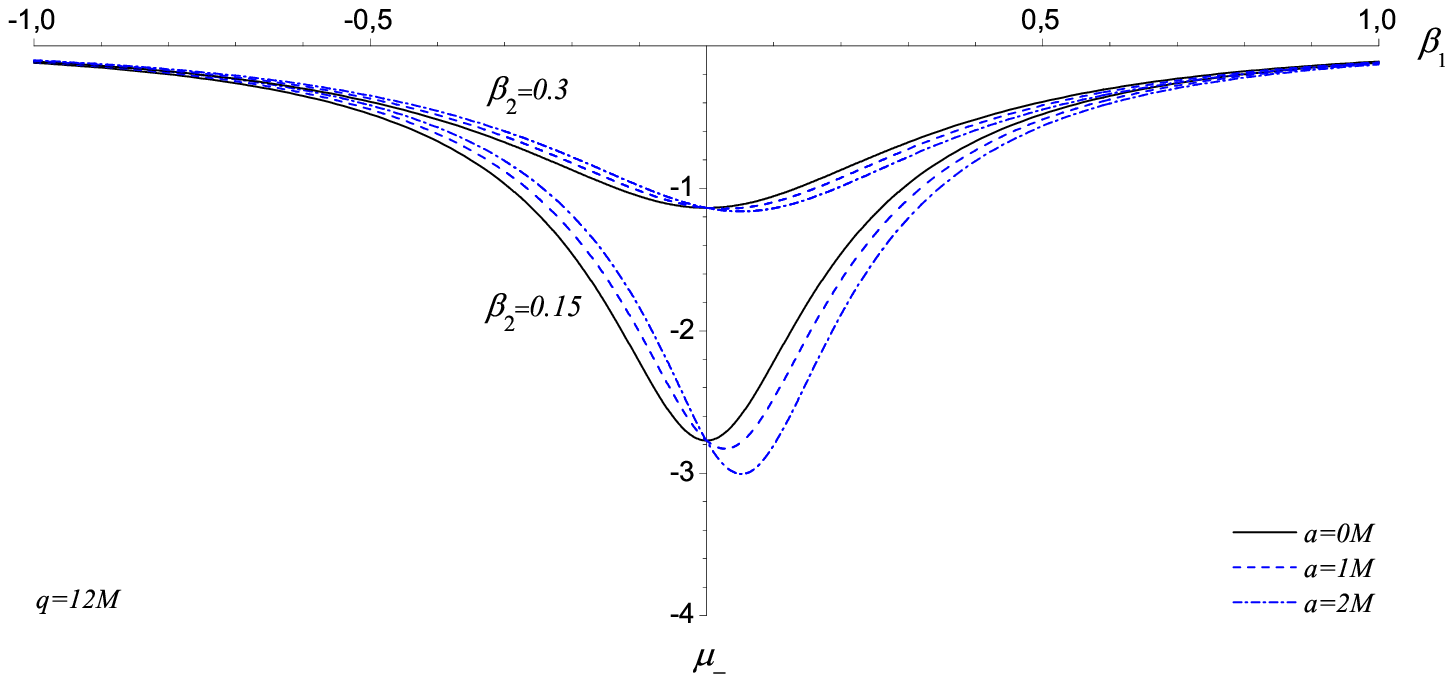} \\
    \caption{\small The magnification of the negative parity image $\mu_{-}$ as a function of the scaled angular coordinate $\beta_{1}$ of the source for different scaled angular coordinate $\beta_{2}$. The Janis--Newman--Winicour naked singularity (solid line) and the rotating naked singularity lenses (dashed line; dashed, doted line) are considered. (a) and (b) are the same as (a) and (b) of Fig. \ref{SM1}.} \label{SM2}
\end{figure}
%----------------------------------------------------------------------------------------
%----------------------------------------------------------------------------------------
\begin{figure}
    \includegraphics[width=0.9\textwidth]{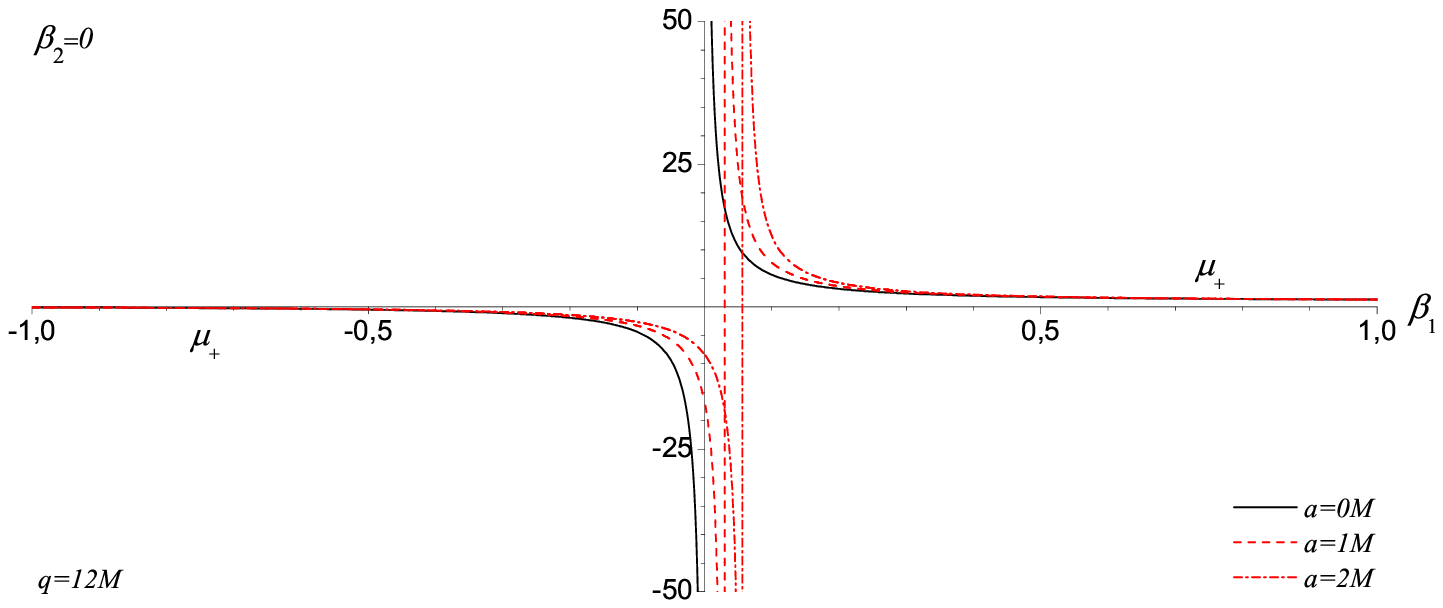} \\
    \vspace{1.0cm}
    \includegraphics[width=0.9\textwidth]{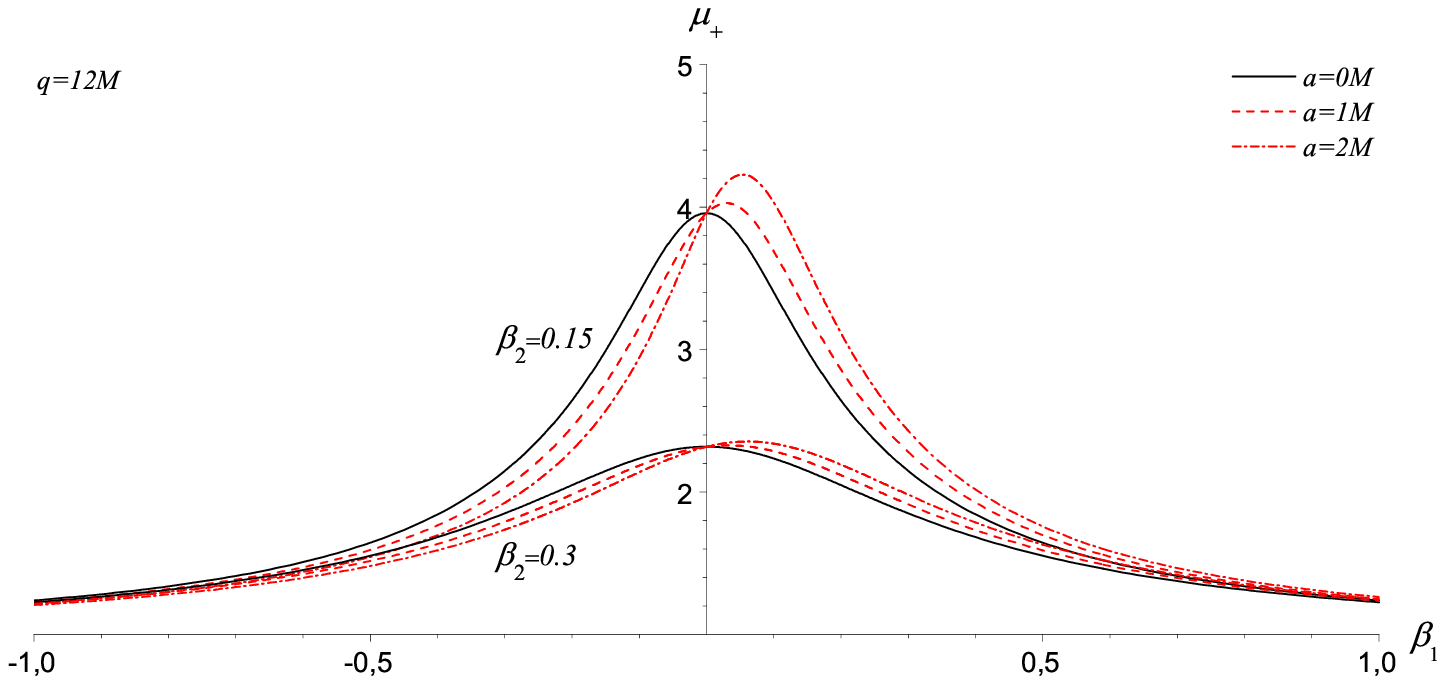} \\
    \caption{\small The magnification of the positive parity $\mu_{+}$ as a function of the scaled angular coordinate $\beta_{1}$ of the source for different scaled angular coordinate $\beta_{2}$. The Janis--Newman--Winicour naked singularity (solid line) and the rotating naked singularity lenses (dashed line; dashed, doted line) are considered. (a) and (b) are the same as (a) and (b) of Fig. \ref{SM1}.} \label{SM3}
\end{figure}
%----------------------------------------------------------------------------------------

In the vicinity of the caustics (or the critical curves) the power series (\ref{SignMagn2}) does not work properly and describe the image magnifications exactly for scaled source angular positions $\beta_{1}\neq\beta_{1}^{cau}$ (or $\theta_{1}\neq\theta^{cr}_{1}$). Therefore, in order to disentangle the influence of the scalar charge and the lens angular momentum over the signed magnifications, in Figs. \ref{SM1}, \ref{SM2} and \ref{SM3} we have plotted the reciprocal values of the Jacobian corresponded to $\mu_{+}$ and $\mu_{-}$ according to the current estimates for the massive dark object in the center of our Galaxy \cite{Eis} as we assume that the observer is equatorial.

In the static case of the Janis--Newman--Winicour lens for all values of the scaled source angular positions $\beta_{1}$ and $\beta_{2}$ the magnification of the positive parity $\mu_{+}$ and the value of that of the negative parity image $\mu_{-}$ increases with the increase of the scalar charge. Let us set the source on the equatorial plane (\textit{i.e.} $\beta_{2}=0$) and move it to the optical axis. Then two weak field images will appear, one on each side of the optical axis. The signed image magnifications start to grow from $\mu_{+}=1$ and $\mu_{-}=0$ respectively for the positive and negative parity images when the absolute value of the source scaled angular coordinate is at infinity. In the case when the source is on the optical axis ($(\beta_{1},\beta_{2})=(0,0)$) the signed magnifications diverge and infinitely bright Einstein rings appear. Removing the source from the optical axis (\textit{i.e.} $\beta_{2}\neq0$) and keeping $\beta_{1}=0$ we will see two weak field images situated respectively on each side of the optical axis in the directions perpendicular to the equatorial plane. Their magnifications decrease with the increase of $\beta_{2}$ and approach the limit values $\mu_{+}=1$ and $\mu_{-}=0$ respectively for the positive and negative parity image when the source scaled angular coordinate goes to infinity. In all of these cases the presence of the scalar charge leads to an increment of the signed magnifications.

When the lens is rotational and the source is equatorial the angular momentum of the lens decreases the signed magnification of the negative parity image and increases the value of that of the positive parity image for every value of $\beta_{1}$ and a scalar charge $q/M$. For the source position $\beta_{1}<\beta_{1}^{cau}$ the negative parity image is outside the critical curve while the positive parity image is inside. Moving the source from infinity to the left hand side of the optical axis the negative parity image magnification starts to grow from $\mu_{-}=1$ while the positive parity image magnification start to decrease from $\mu_{+}=0$. Passing through the optical axis the source reaches the point-like caustic $(\beta_{1}^{cau},0)=({a}/(\theta_{E}D_{OL})+\mathcal{O}(\epsilon^3),0)$, then the value of the signed image magnifications diverges and infinitely bright critical curves appear. Drawing back the equatorial source from the left hand side of the caustic point the negative signed magnification increases, while the positive signed magnification decreases with the increase of $\beta_{1}$ and approaches respectively $\mu_{-}=0$ and $\mu_{+}=1$ when $\beta_{1}\rightarrow\infty$. When we remove the source from the equatorial plane for every $\beta_{2}$ and $\beta_{1}<0$ the lens angular momentum leads to an increase in the negative parity magnification and to a decrease in the positive parity magnification in comparison to the magnification of the static case. When $\beta_{1}>0$ the negative parity magnification decreases, while the positive parity magnification increases. The signed magnifications coincide for a source in the point ($0,\beta_{2}$) where the static signed magnifications have a maximum. When $\beta_{2}\neq0$ non-equatorial motion of the source leads to occurrence of the positive parity image outside of the critical curve and to raising of the negative parity image inside the critical curve. For fixed value of scaled angular coordinate $\beta_{1}$ and lens angular momentum $a$ as well as all values of $\beta_{2}$ the values of the positive and the negative parity image magnifications increase with the increase of the scalar charge $q/M$.

For big enough values of the scalar charge $q/M$ and the lens angular momentum $a$ the graphics do not describe reliable authenticity of the pictured quantities in the post-Newtonian order, which works sufficiently well for small values of the source scaled angular coordinates $\beta_{1}$ and $\beta_{2}$. Therefore, we do not show this situations.

\subsection{Total magnification and centroid.}

When the two weak field images are not resolved and are packed together the main observables become the total magnification and magnification-weighted centroid position. Taking into account that the image parities give for to the absolute magnifications $|\mu^{+}|=\mu^{+}$ and $|\mu^{-}|=-\mu^{-}$, then the total absolute magnification for the rotating singularities space-time is
\begin{equation}\label{MangTot}
    \mu_{tot}=|\mu^{+}|+|\mu^{-}|=\frac{2+\beta^2}{\beta\sqrt{4+\beta^2}}+\frac{8\beta_{1}}{\beta^{3}(4+\beta^2)^{3/2}}\frac{a\sin\vartheta_{O}}{M}\epsilon+\mathcal{O}(\epsilon^2),
\end{equation}
and up to terms $\mathcal{O}(\epsilon^2)$ do not differ from the result for the Kerr lensing \cite{WerPet}. When the observer is on the rotational axis or in the particular case of the circularly symmetric lens ($a=0$), the term $\mathcal{O}(\epsilon)$ vanishes.
%----------------------------------------------------------------------------------------
\begin{figure}
    \includegraphics[width=0.99\textwidth, height=0.29\textheight]{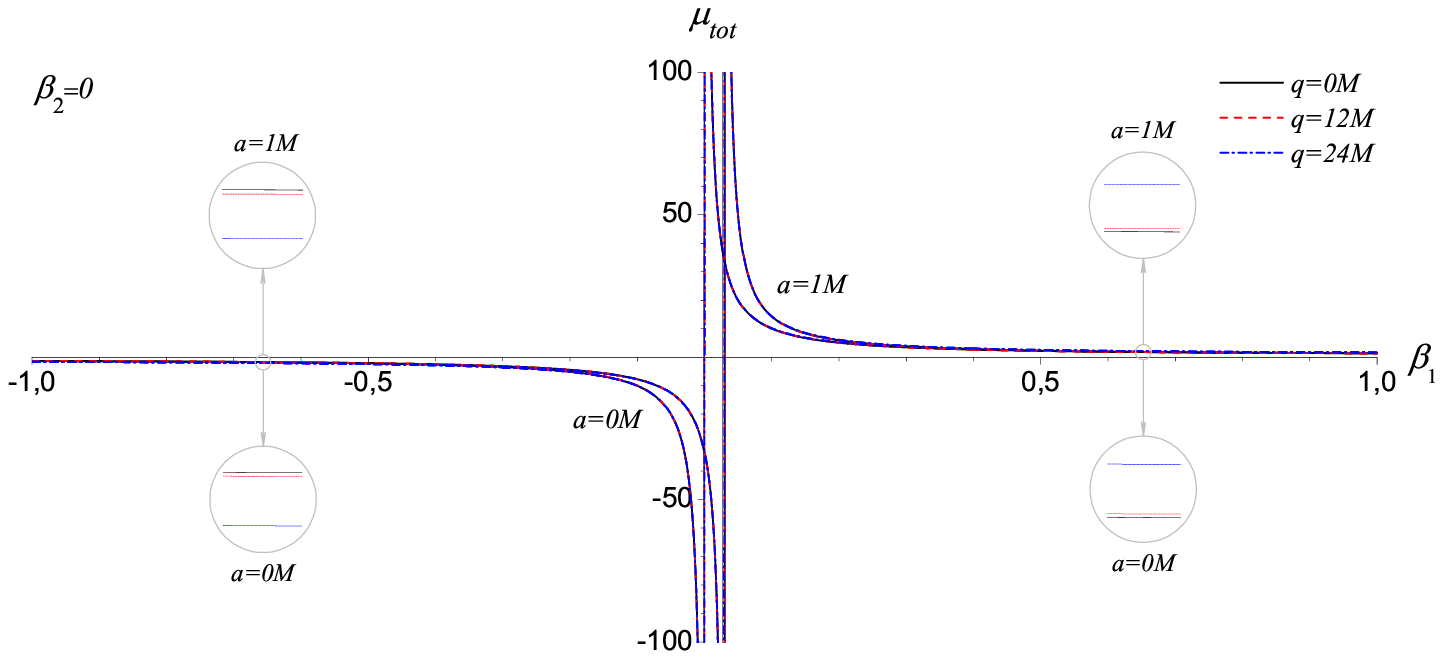} \\
    \vspace{0.5cm}
    \includegraphics[width=0.99\textwidth, height=0.29\textheight]{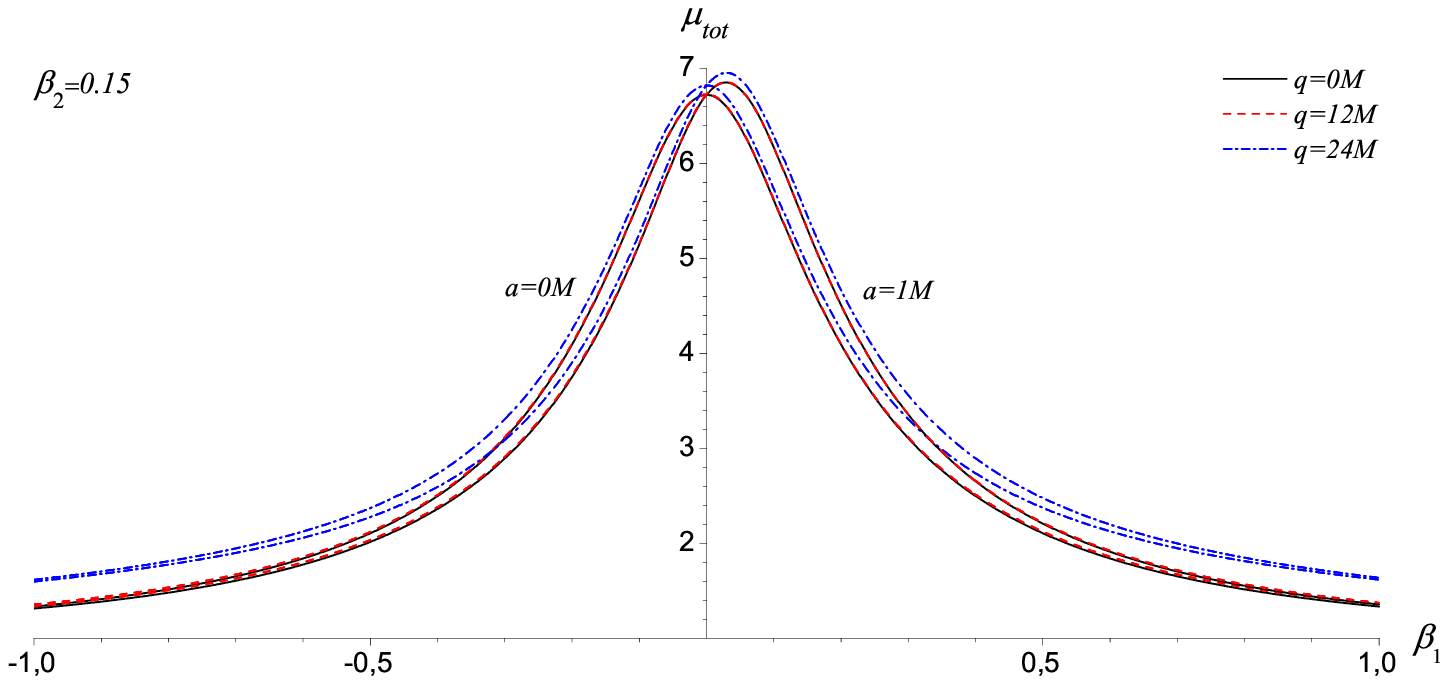} \\
    \vspace{0.5cm}
    \includegraphics[width=0.99\textwidth, height=0.29\textheight]{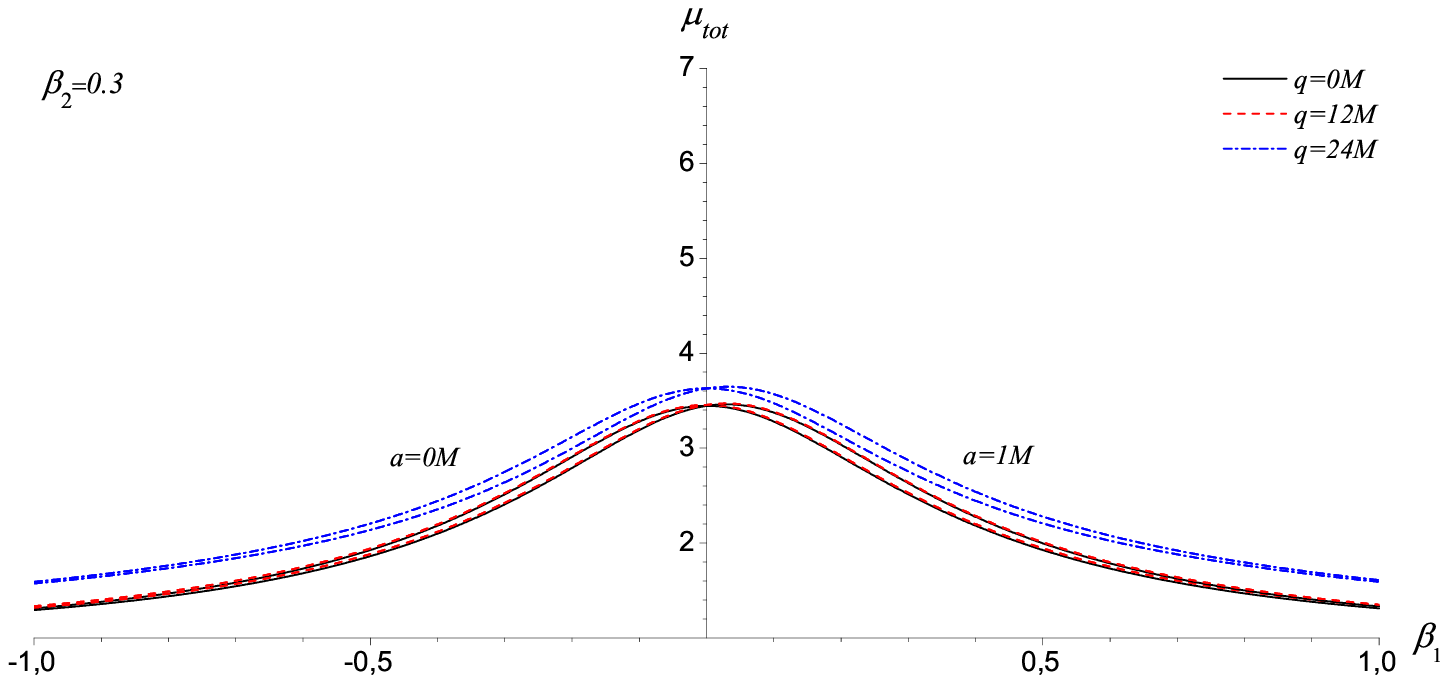}
    \caption{\small The total magnification $\mu_{tot}$ as a function of the scaled angular coordinate $\beta_{1}$ of the source for different scaled angular coordinate $\beta_{2}$. The Schwarzschild ($a=0M$ and $q=0M$) and the extremal Kerr black hole lens ($a=1M$ and $q=0M$) as well as the Janis--Newman--Winicour ($a=0M$ and $q=12M$, $24M$) and the rotating naked singularity lenses ($a=1M$ and $q=12M$, $24M$) are considered. (a) and (b) are the same as (a) and (b) of Fig. \ref{SM1}.} \label{TotMagn}
\end{figure}

%----------------------------------------------------------------------------------------
%----------------------------------------------------------------------------------------
\begin{figure}
    \includegraphics[width=0.465\textwidth]{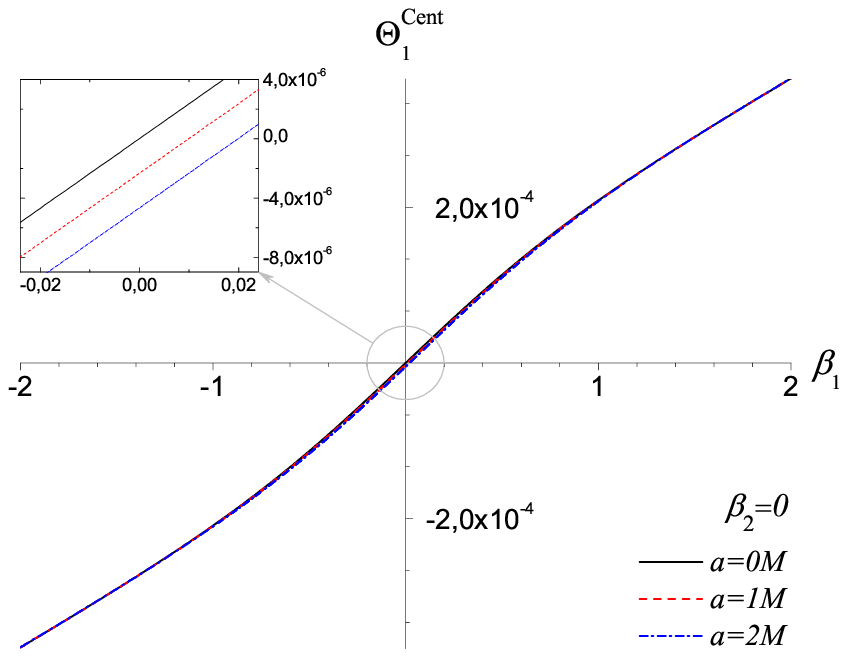} \hspace{0.3cm}
    \includegraphics[width=0.465\textwidth]{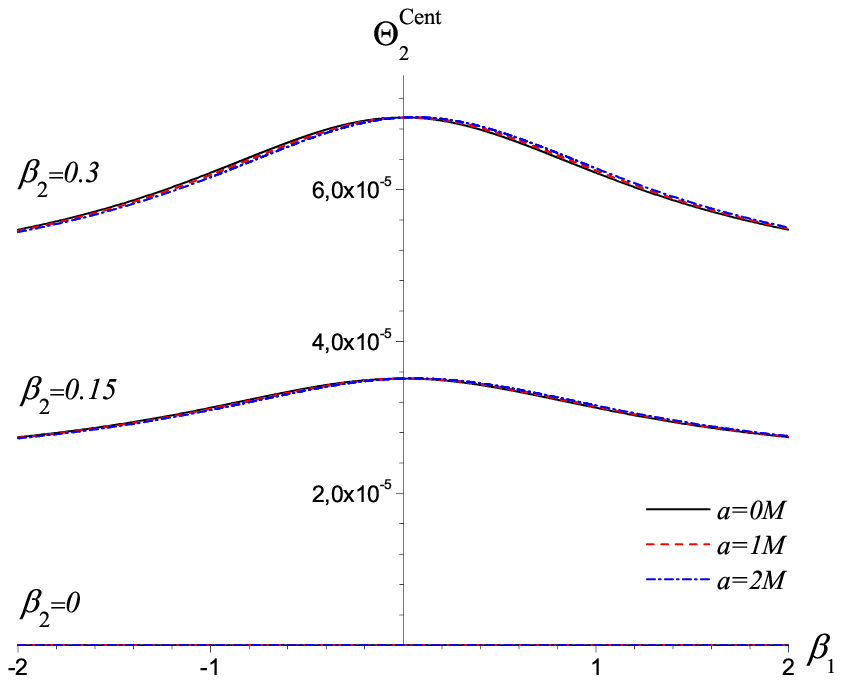} \\
    \caption{\small The magnification centroid components $\Theta_{1}^{\rm Cent}$ and $\Theta_{2}^{\rm Cent}$ as a function of the scaled angular coordinate $\beta_{1}$ of the source for different scaled angular coordinate $\beta_{2}$. The Schwarzschild black hole (solid line) and the extremal Kerr black hole (dashed line) as well as the Kerr naked singularity lenses (dashed, doted line) lens are considered. (a) and (b) are the same as (a) and (b) of Fig. \ref{SM1}. (b) $\Theta_{1}^{\rm Cent}$ and $\Theta_{2}^{\rm Cent}$ are expressed in arcseconds. } \label{Cent12}

\hspace{2.0cm}
%\end{figure}
%----------------------------------------------------------------------------------------
%\begin{figure}
    \includegraphics[width=0.99\textwidth]{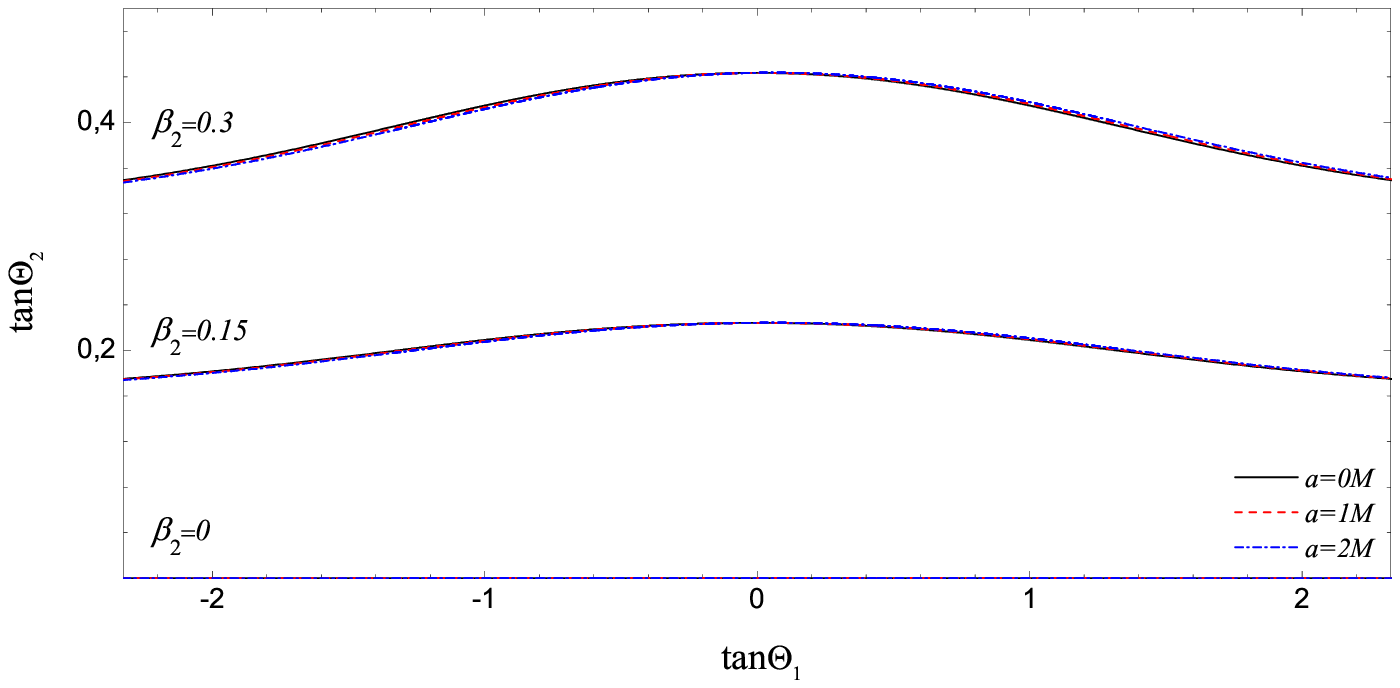}
    \caption{\small The magnification centroid $\Theta^{\rm Cent}$ is plotted in the lane $\{\tan\Theta_{1},\tan\Theta_{2}\}$ for source scaled angular coordinate $\beta_{1}\in[-2$, $2]$ and different scaled angular coordinate $\beta_{2}$. The Schwarzschild black hole (solid line) and the extremal Kerr black hole (dashed line) as well as the Kerr naked singularity (dashed, doted line) lenses are considered. (a) is the same as (a) of Fig. \ref{SM1}. (b) Axis-lengths are in units of tangent of Einstein angle, $\theta_{E}\simeq157$ $\mu$arcsec.} \label{Cent}
\end{figure}
%----------------------------------------------------------------------------------------
The magnification centroid position is defined by \cite{GauPett}
\begin{equation}\label{Centroid}
    \boldsymbol{\Theta}^{\rm Cent}=\frac{\boldsymbol{\theta^{+}}|\mu^{+}|+\boldsymbol{\theta^{-}}|\mu^{-}|}{|\mu^{+}|+|\mu^{-}|}.
\end{equation}
Using (\ref{Theta_Sch}), (\ref{Theta_11}) and (\ref{MuPlasMinus}) we obtain the same expressions
as those for Kerr black hole lensing up to post-Newtonian order \cite{WerPet}
\begin{eqnarray}
  \Theta_{1}^{\rm Cent} &=& \theta_{E}\left[\frac{(3+\beta^{2})\beta_{1}}{2+\beta^{2}}+\frac{(\beta_{1}^2-\beta_{2}^2-2)}{(2+\beta^2)^2}\frac{a\sin\vartheta_{O}}{M}\epsilon+\mathcal{O}(\epsilon^2)\right], \nonumber \\
  \Theta_{2}^{\rm Cent} &=& \theta_{E}\left[\frac{(3+\beta^{2})\beta_{2}}{2+\beta^{2}}+\frac{2\beta_{1}\beta_{2}}{(2+\beta^2)^2}\frac{a\sin\vartheta_{O}}{M}\epsilon+\mathcal{O}(\epsilon^2)\right].  \end{eqnarray}
For the case $a=0$ or $\vartheta_{O}=0$, the magnification centroid for the rotating singularity space-time coincides with the result by Keeton and Petters \cite{KeePett}.

Since the total magnification is the difference of the reciprocal values of the Jacobi determinant related to the positive and the negative parity images, we have plotted it in Fig. \ref{TotMagn} as a function of the source scaled angular coordinates $\beta_{1}$ and $\beta_{2}$ for different values of the lens angular momentum and the scalar charge. The magnification centroid is plotted also in Figs. \ref{Cent12} and \ref{Cent}. The total magnification has a similar behavior as the positive parity magnification, with the difference that for a fixed $a$ with the increase of the scalar charge $\mu_{tot}$ increases for $\beta_{1}>0$ and decreases for $\beta_{1}<0$. For the equatorial source (\textit{i.e.} $\beta_{2}=0$) the magnification centroid $\Theta^{\rm Cent}=\Theta_{1}^{\rm Cent}$ grows with an increase of $\beta_{1}$. When the source is near the optical axis the lens angular momentum decreases more the magnification centroid and less for bigger values of the source scaled angular coordinate $\beta_{1}$, where the graphics coincide and become straight lines. When the source drifts away from the equatorial plane (\textit{i.e.} $\beta_{2}\neq0$) the dependence of $\Theta_{1}^{\rm Cent}$ on $a$ decreases for bigger values of $\beta_{2}$. The second magnification centroid coordinate $\Theta_{2}^{\rm Cent}$ increases when $\beta_{2}$ grows and has a maximum for $\beta_{1}=0$. As lens the angular momentum $a$ increases $\Theta_{2}$ first decreases for the source on the left hand side of the optical axis and then increases for a source on the opposite side of the axis. When the source is at the point ($0,\beta_{2}$) the magnification centroid curves $\Theta_{2}$ coincide with those for the static case. The magnification centroid $\Theta^{\rm Cent}$ is plotted in the Fig. \ref{Cent} over the image plane $\{\tan{\Theta_{1}}, \tan{\Theta_{2}}\}$. For different values of the lens parameters $a$ and $q/M$ and also for the scaled angular coordinates of the source $\beta_{1}$ and $\beta_{2}$ the graphics resemble the behavior of $\Theta_{2}^{\rm Cent}$.

\section{Exact numerical investigation of the deflection angle}

Let us impose the condition $\theta=\pi/2$ and set the light ray on the equatorial plane.
In this case, if we substitute $x=r/2M$ as a new radial coordinate and measure all distances in units $2M=1$ we obtain the reduced
metric in the form
\begin{eqnarray}
  ds^{2}=A(x)dt^{2}-B(x)dx^{2}-C(x)d\phi^{2}+D(x)dtd\phi. \label{RMet}
\end{eqnarray}
The metric components are:
\begin{eqnarray}
 && A(x)=\left(1-\frac{1}{\gamma{x}}\right)^{\gamma}, \\
 && B(x)=\frac{x^{2}}{[x^{2}+a^{2}-\frac{x}{\gamma}]}\left(1-\frac{1}{\gamma{x}}\right)^{1-\gamma}, \\
 && C(x)=a^{2}\left[2-\left(1-\frac{1}{\gamma{x}}\right)^{\gamma}\right]+x^{2}\left(1-\frac{1}{\gamma{x}}\right)^{1-\gamma},  \\
 && D(x)=2a\left[1-\left(1-\frac{1}{\gamma{x}}\right)^{\gamma}\right].
\end{eqnarray}

The relation between the impact parameter (the perpendicular distance from the lens to
the tangent to the null geodesic of the source) and the distance of closest approach of the light ray $x_{0}$ can be obtained
from the conservation of the angular momentum of the scattering process, and it is given by
\begin{equation}\label{ImpactParameter}
   J(x_{0})=\frac{a\left(1-\frac{1}{\gamma{x}_{0}}\right)^{\gamma}-a+\sqrt{x_{0}^{2}+a^{2}-\frac{x_{0}}{\gamma}}}{\left(1-\frac{1}{\gamma{x}_{0}}\right)^{\gamma}}. \end{equation}
The sign in front of the square root is chosen to be
positive when the light ray is winding counterclockwise. For $a>0$
the black hole rotates counterclockwise, while for $a<0$ the black
hole and the photons rotate in converse direction.
%------------------------------------------------------------------------------------------------------------------------
\begin{figure}
  \includegraphics[width=0.5\textwidth]{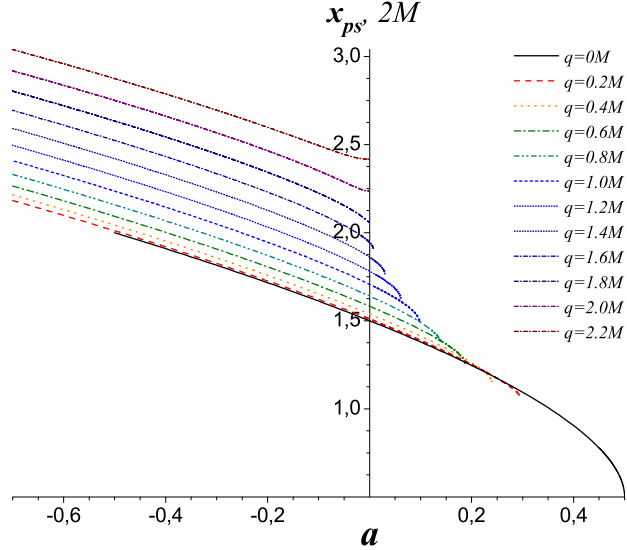}\\
  \caption{\small The radius of the photon sphere as a function of the lens angular momentum $a$ for a scalar charge  $q/M=0$ (solid line) to a scalar charge $q/M=2.2$ (short, dashed, dotted line) by $q/M=0.2$.}  \label{FS}
\end{figure}
%------------------------------------------------------------------------------------------------------------------------

A starting point of the strong field limit expansion is the photon
sphere, which has been defined by Virbhadra and Ellis \cite{Vir1}
(and has subsequently been investigated by Claudel, Virbhadra and
Ellis \cite{PSphere}) as a time-like hypersurface $\{{x}=x_{ps}\}$ on
which the light bending angle becomes unboundedly large when the
closest distance of approach $x_{0}$ tends to $x_{ps}$. The photon
sphere equation for a stationary, axially-symmetric metric
is
\begin{equation}
   A_{0}C_{0}^{\prime}-A_{0}^{\prime}C_{0}+J(A_{0}^{\prime}D_{0}-A_{0}D_{0}^{\prime})=0, \end{equation}
for which we require to admit at least one positive solution. For the Kerr-like metric the photon sphere equation takes the form
\begin{equation}\label{FS_AE}
    4\gamma^{3}x^{4}-8(1+\gamma)\gamma^{2}x^{3}+\gamma(1+2\gamma)(5+2\gamma)x^{2}-[(1+2\gamma)^{2}+8a^{2}\gamma^{3}]x+4a^{2}\gamma^{2}(1+\gamma)=0.
\end{equation}
The biggest real root of this equation external to the outer Kerr black hole horizon or to the curvature singularity
 defines the radius of the photon sphere $x_{ps}=r_{ps}/2M$. The radius of the photon sphere is computed numerically
and is plotted in Fig. \ref{FS} for different values of the ratio
$q/M$. The analytical investigation of the fourth order algebraic
Eq. (\ref{FS_AE}) shows the kind of the naked singularity.
Depending on whether or not the naked singularity is
covered within a photon sphere Virbhadra and Ellis \cite{Vir2}
classify the naked singularity as \textit{weakly naked} (WNS) (those
contained within at least one photon sphere) and \textit{strongly
naked} (SNS) (those not contained within any photon sphere). In the
particular case of $\gamma=1$ when  the Kerr black hole is
recovered, the photon sphere exists for angular momentum
$-1\leq{2a}\leq1$. In the case of rotating naked singularities when
$1/2<\gamma<1$ the photon sphere exists for angular momentum
$-\infty<a\leq{a}_{m}$, while for $0<\gamma\leq1/2$ the naked
singularity allows a photon sphere for angular momentum
$-\infty<a<0$.  Here $a_{m}$ is defined by
\begin{equation}
    a_{m}=\frac{\sqrt{2\gamma(9\gamma-\sqrt{45\gamma^{2}-192\gamma^{6}+144\gamma^{4}+3})}}{12\gamma^{2}}.
\end{equation}
An intermediate position takes the static case $\gamma=1/2$ and $a=0$, when the so called \textit{marginally strongly naked singularity} (MSNS) is realized. In this case there is no photon sphere, but the gravitational lensing leads to the appearing of relativistic images. We do not show the numerically investigation to this situation, which is qualitatively similar to the static WNS.

The Kerr black hole and WNS photon spheres decrease with the increase
in $|a|$ and have a similar behavior for different scalar charges.
In the case of the Kerr black hole ($q/M=0$) when $a$ reaches its extremal value, the photon sphere coincides with the
black hole horizon. With the increase of $q/M$ the
photon sphere $x_{ps}$ and the curvature singularity $x_{cs}$
increase as $x_{ps}>x_{cs}$ for $a\leq{a}_{m}$ when the
singularities are weakly naked and meet at $(q/M)^{2}=3$ and $a=0$.
Moreover, for the bigger values of the scalar charge the photon
sphere $x_{ps}\equiv{x}_{cs}$ in the static case $a=0$ and therefor
in this cases the singularity is strongly naked. The
different values of the upper limit of WNS angular momentum, for
which a photon sphere exists and the
limits of the photon sphere corresponding to them
can be seen on Fig. \ref{FS} and are also written in Table
\ref{UpperAngularMomentum}.
%----------------------------------------------------------------------------------------------------------------------
\begin{table}[t]
\caption{Upper limit of the lens angular momentum still holding up a photon sphere.}
\centering
\begin{tabular}{c c c c c c c c c c c c c}
\hline\hline
 & Kerr BH & \multicolumn{11}{c}{WNS} \\ [0.5ex]
\hline
$q/M$ & 0 & 0.2 & 0.4 & 0.6 & 0.8 & 1.0 & 1.2 & 1.4 & 1.6 & 1.8 & 2.0 & 2.2 \\
$\gamma$ & 1 & 0.981 & 0.928 & 0.857 & 0.781 & 0.707 & 0.640 & 0.581 & 0.530 & 0.486 & 0.447 & 0.414 \\
$a_{max}/2M$ & 0.5 & 0.294 & 0.238 & 0.187 & 0.139 & 0.097 & 0.060 & 0.029 & 0.007 & $\rightarrow0$ & $\rightarrow0$ & $\rightarrow0$ \\
$\lim_{a\rightarrow{a_{max}}}x_{ps}$ & 0.5 & 1.068 & 1.154 & 1.256 & 1.371 & 1.496 & 1.627 & 1.764 & 1.905 & 2.059 & 2.236 & 2.417\\[1ex]
\hline\hline
\end{tabular}
\label{UpperAngularMomentum}
\end{table}
%----------------------------------------------------------------------------------------------------------------------

The bending angle of a light ray in a stationary, axially-symmetric space-time, described by the line element (\ref{RMet}) is given by
\begin{eqnarray}\label{Integral}
&&\tilde{\alpha}(x_{0})=\phi_{f}(x_{0})-\pi, \nonumber \\
&&\phi_{f}(x_{0})=2\int_{x_{0}}^{\infty}\frac{d\phi}{dx}dx, \nonumber \\
&& \frac{d\phi}{dx}=\pm\frac{\sqrt{B|A_{0}|}(D+2JA)}{\sqrt{C}\sqrt{4AC+D^{2}}\sqrt{sgn(A_{0})[A_{0}-A\frac{C_{0}}{C}+\frac{J}{C}(AD_{0}-A_{0}D)]}},
\end{eqnarray}
where $\phi_{f}(x_{0})$ is the total azimuthal angle. With the
decrease of the distance of closest approach $x_{0}$ the
deflection angle increases, and for a certain value of $x_{0}$ the
deflection angle becomes $2\pi$, so that the light makes a complete
loop around the lens. Let $x_{0}$ decreases further, then the light
ray will wind several times around the lens before reaching the
observer and finally when $x_{0}$ becomes equal to the radius of the
photon sphere $x_{ps}$ the deflection angle will become unboundedly
large and the photon will be captured by the lens object.
%------------------------------------------------------------------------------------------------------------------------
\begin{figure}
\includegraphics[width=0.32\textwidth,height=0.25\textwidth]{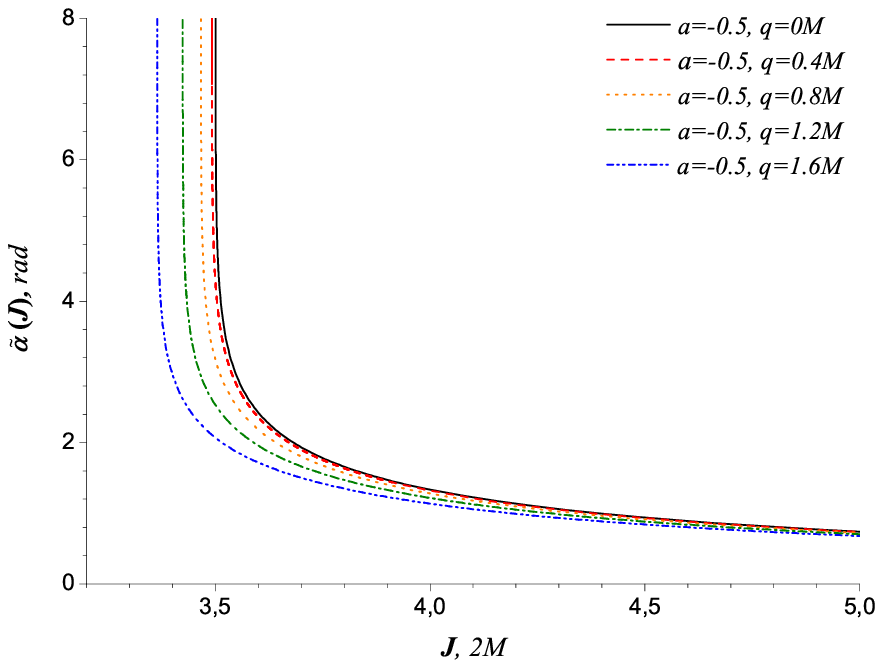}
\includegraphics[width=0.32\textwidth,height=0.25\textwidth]{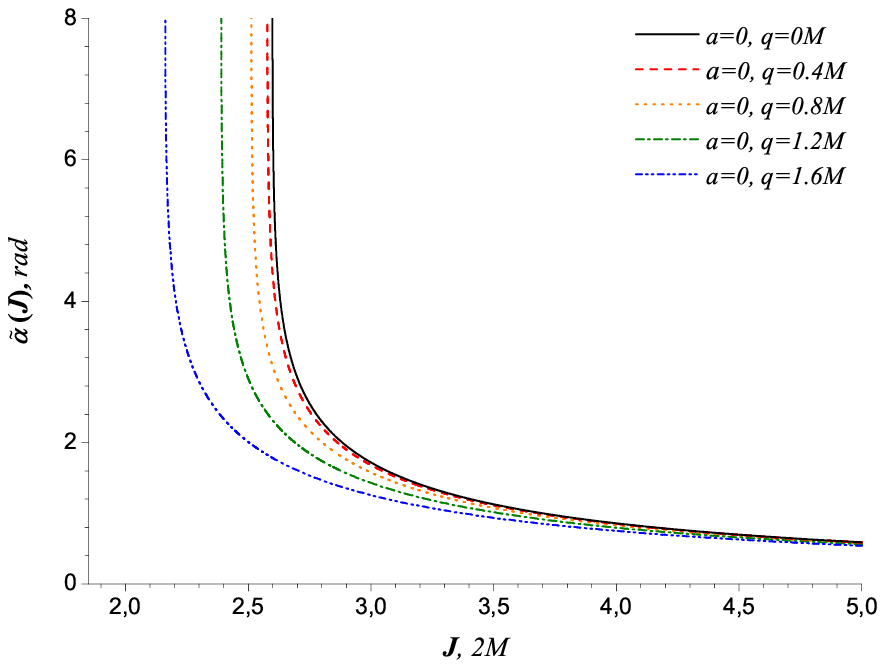}
\includegraphics[width=0.32\textwidth,height=0.25\textwidth]{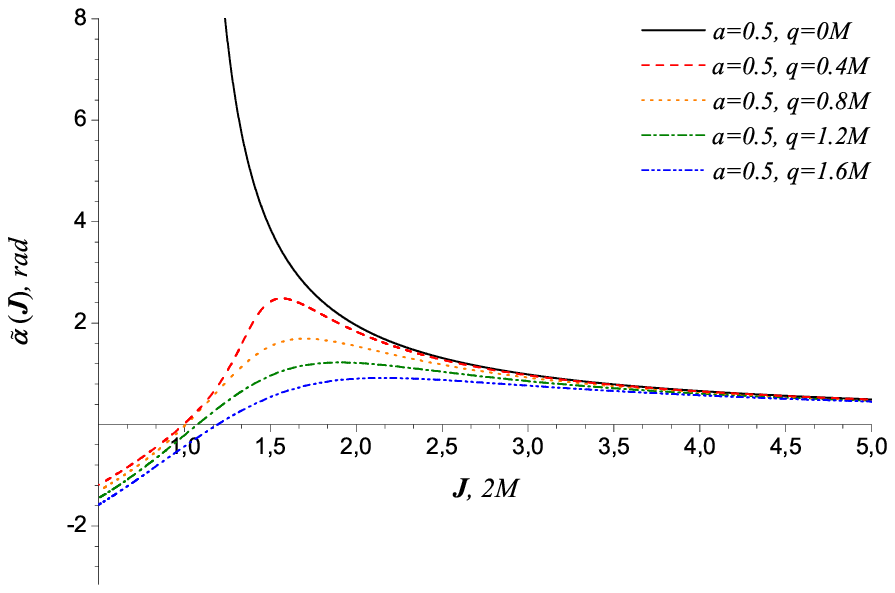}\\
\vspace{0.4cm}
\includegraphics[width=0.32\textwidth,height=0.25\textwidth]{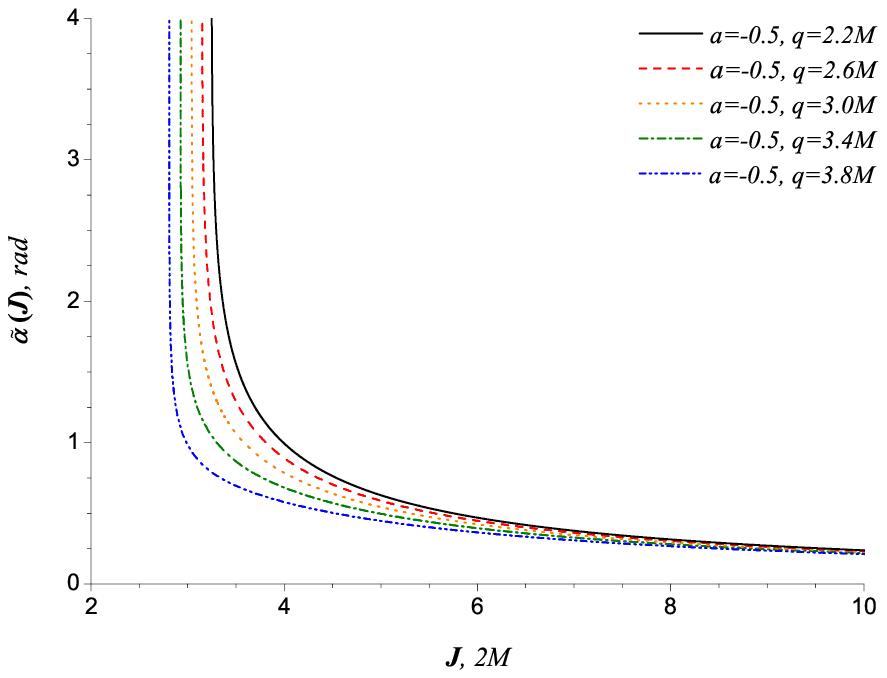}
\includegraphics[width=0.32\textwidth,height=0.25\textwidth]{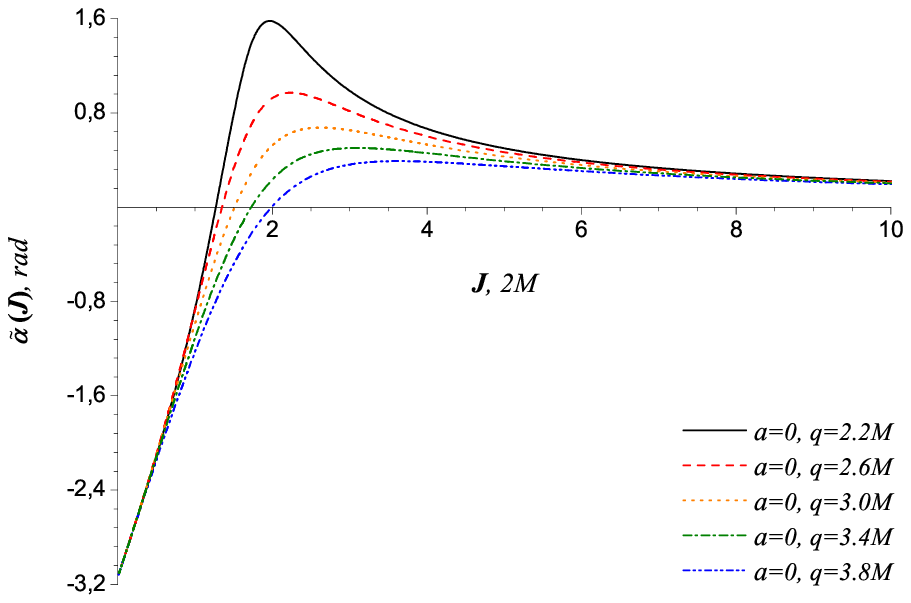}
\includegraphics[width=0.32\textwidth,height=0.25\textwidth]{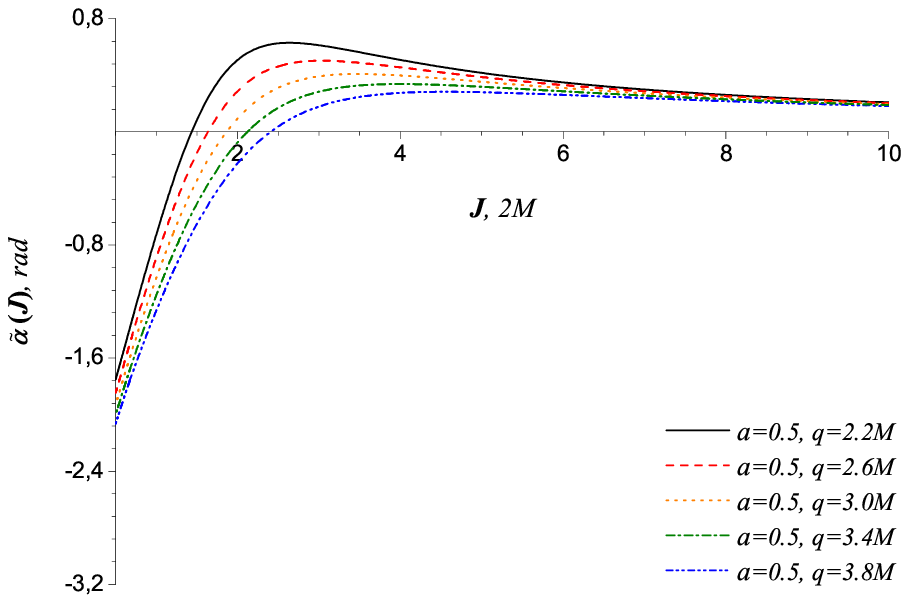}\\
\vspace{0.4cm}
\includegraphics[width=0.32\textwidth,height=0.25\textwidth]{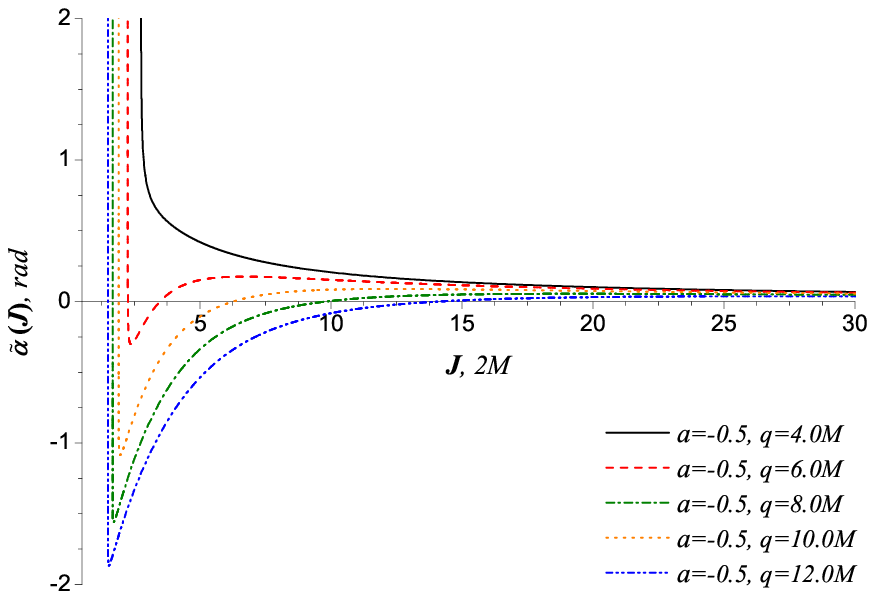}
\includegraphics[width=0.32\textwidth,height=0.25\textwidth]{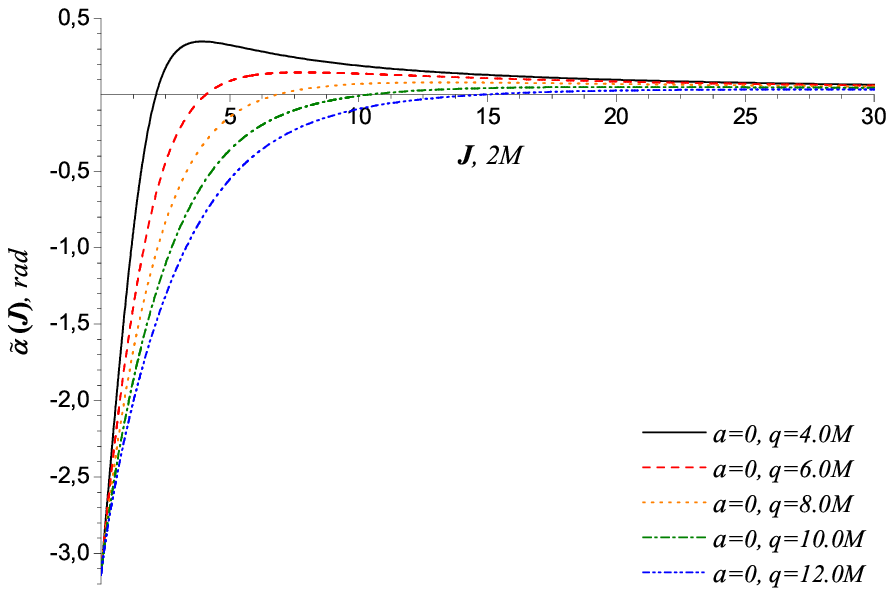}
\includegraphics[width=0.32\textwidth,height=0.25\textwidth]{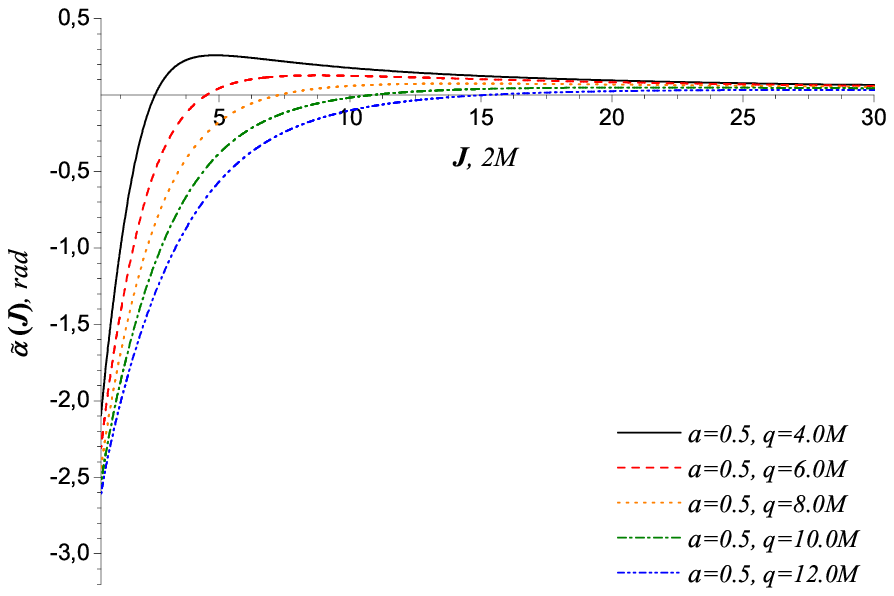}
\caption{\small Deflection angle $\tilde{\alpha}(J)$ as a function
of impact parameter $J$ for angular momentum $a=-0.5$ (left column),
$a=0$ (central column) and $a=0.5$ (right column) and for
different scalar charges $q/M=0\div1.6$ by charge $q/M=0.4$ (first
row), $q/M=2.2\div3.8$ by charge $q/M=0.4$ (second row) and
$q/M=4\div12$ by charge $q/M=2.0$ (third row).}\label{DAngles}
\end{figure}
%------------------------------------------------------------------------------------------------------------------------
Let us discuss the behavior of the bending angle $\tilde{\alpha}$ for the Kerr black hole, WNS and SNS. One has
$\lim_{x_{0}\rightarrow\infty}\tilde{\alpha}(x_{0})=0$ for all values of $\gamma$ and the angular momentum $a$,
$\lim_{x_{0}\rightarrow{x}_{ps}}\tilde{\alpha}(x_{0})=\infty$ for $\gamma=1$, $-1\leq{2a}\leq1$ (Kerr black hole) and $0<\gamma\leq1/2$,
$-\infty<a<0$ and $1/2<\gamma<1$, $-\infty<a\leq{a}_{m}$ (weakly naked singularities) and $\lim_{x_{0}\rightarrow{x}_{cs}}\tilde{\alpha}(x_{0})=
\tilde{\alpha}_{min}$ for $0<\gamma\leq1/2$, $0\leq{a}<\infty$ and $1/2<\gamma<1$, $a_{m}<a<\infty$ (strongly naked
singularities). $\tilde{\alpha}_{min}$ is the minimum of the deflection angle for SNS, which is a function of the angular
momentum $a$ and the scalar charge $q/M$ and is given by
\begin{eqnarray}
\lim_{x_{0}\rightarrow{x}_{cs}}\tilde{\alpha}(x_{0})&=&-\pi+\frac{2a\gamma}{\sqrt{1-4a^2\gamma^2}}\ln\bigg{|}\frac{1+\sqrt{1-4a^2\gamma^2}}{1-\sqrt{1-4a^2\gamma^2}}\bigg{|}, \,\,\,\,\,\,\,\,\,\,\,\,\,\,\,\,\,\,\,\,\,\,\,\,\, 4a^2\gamma^2<1,\\
    &=&-\pi+4a\gamma,\,\,\,\,\,\,\,\,\,\,\,\,\,\,\,\,\,\,\,\,\,\,\,\,\,\,\,\,\,\,\,\,\,\,\,\,\,\,\,\,\,\,\,\,\,\,\,\,\,\,\,\,\,\,\,\,\,\,\,\,\,\,\,\,\,\,\,\,\,\,\,\,\,\,\,\,\,\,\,\,\,\,\,\,\,\,\,\,\,\,\,\,\,\,\,\,\,\,\,\, 4a^2\gamma^2=1,\\
    &=&-\pi+\frac{2a\gamma}{\sqrt{4a^2\gamma^2-1}}\bigg{[}\pi-2\arctan{\frac{1}{\sqrt{4a^2\gamma^2-1}}}\bigg{]}, \,\,\,\,\,\,\,\, 4a^2\gamma^2>1.
\end{eqnarray}

In Fig. \ref{DAngles} we plot the behavior of the bending angle
$\tilde{\alpha}$ against the impact parameter $J$ (scaled in terms
of $2M$) for different values of the scalar charges $q/M$ and the
lens angular momentum $a$. In the top left, top middle and middle left figures we
plot the curves respectively for $a=-0.5$ and a scalar charge
$q/M=0\div1.6$ (WNS) with a step $q/M=0.4$, $a=0$ and a scalar
charge $q/M=0\div1.6$ (WNS) with a step $q/M=0.4$ and for $a=-0.5$
and a scalar charge $q/M=2.2\div3.8$ (WNS) with a step $q/M=0.4$. It
is obvious that they have similar qualitative behavior for the
different scalar charges; the bending angle strictly increases with
the decreases of the impact parameter and becomes unboundedly large
as the impact parameter approaches the respective value it
has on the photon spheres (i. e $J\rightarrow J_{ps}$). In Fig.
\ref{DAngles} (bottom left) we plot the curves for
$a=-0.5$ and scalar charge $q/M=4\div12$ (WNS) with a step
$q/M=2.0$; for all curves but the uppermost
($q/M=4$) the deflection angle changes its sign as it first
increases up to its maximum with a decrease in the impact parameter,
after that becomes null with the further decreases of the impact
parameter and finally, after reaching its minimum
$\tilde{\alpha}$ becomes unboundedly large as $J\rightarrow J_{ps}$.
In Fig. \ref{DAngles} (top right) we plot curves for positive
angular momentum $a=0.5$ and a scalar charge $q/M=0\div1.6$ with a
step $q/M=0.4$ and without the uppermost (WNS) these are for SNS.
$\tilde{\alpha}$ first increases with the decrease of the
impact parameter and further decreases to the minimum value
${\tilde{\alpha}}_{min}$ as the impact parameter approaches the
impact parameter for their respective $x_{cs}$ (i. e.
$J\rightarrow{J_{cs}}$). As there are no photon spheres for SNS the
deflection angle for these cases are never unboundedly large.
Therefore the gravitational lensing by SNS would not give rise to
relativistic images. In Fig. \ref{DAngles} (center, middle right,
bottom middle, bottom right) we plot for some other values of the
angular momentum $a$ and the scalar charge $q/M$ for SNS.
$\tilde{\alpha}$ translates its maximum to the right
at the bigger impact parameters and decreases with
the increase of the angular momentum $a$ or of the scalar
charge $q/M$.

\section{Gravitational lensing by Kerr black hole and rotating weakly naked singularities in strong deflection limit}

In this section we study the gravitational lensing by Kerr black hole and weakly naked singularities which have a photon sphere. We find the logarithmic behavior of the deflection angle around the photon sphere in the space-time under consideration and discuss the lensing observables.

\subsection{Deflection angle in the stong deflection limit}

Considering equatorial light ray trajectory we can find the behavior of the deflection angle very close to the
photon sphere following the evaluation technique for the integral
(\ref{Integral}) developed by Bozza \cite{Bozza1}. The divergent
integral is first split into two parts one of which
$\phi_{f}^{D}(x_{0})$ contains the divergence and the other
$\phi_{f}^{R}(x_{0})$ is regular. Both pieces are expanded around
$x_{0}=x_{ps}$ and with sufficient accuracy are
approximated with the leading terms. At first, we express the
integrand of (\ref{Integral}) as a function of two the new
variables $y$ and $z$ which are defined by
\begin{eqnarray}
  &&y=A(x),\\
  &&z=\frac{y-y_{0}}{1-y_{0}},
\end{eqnarray}
where $y_{0}=A_{0}$.

 The whole azimuthal angle then takes the form
\begin{equation}
  \phi_{f}(x_{0}) = \int_{0}^{1}R(z,x_{0})f(z,x_{0})dz, \label{AzimEngle}
\end{equation}
where the functions are defined as follows
\begin{eqnarray}
  &&R(z,x_{0})=2\frac{(1-y_{0})}{A^{\prime}}\frac{\sqrt{B|A_{0}|}(D+2JA)}{\sqrt{C}\sqrt{4AC+D^{2}}}, \label{FuncR} \\
  &&f(z,x_{0})=\frac{1}{\sqrt{sgn(A_{0})[A_{0}-A\frac{C_{0}}{C}+\frac{J}{C}(AD_{0}-A_{0}D)]}}. \label{Funcf}
\end{eqnarray}
All functions in (\ref{FuncR}) and (\ref{Funcf}) without the
subscript $''0''$ are evaluated at $ x=A^{-1}[(1-y_{0})z+y_{0}]$.
The function $R(z,x_{0})$ is regular for all values of $z$ and
$x_{0}$ but $f(z,x_{0})$ diverges when $z\rightarrow0$, \textit{i.
e.} as one approaches the photon sphere. The integral
(\ref{AzimEngle}) is then separated in two parts
\begin{equation}
    \phi_{f}(x_{0})=\phi_{f}^{D}(x_{0})+\phi_{f}^{R}(x_{0}),  \end{equation}
where
\begin{equation}
    \phi_{f}^{D}(x_{0})=\int_{0}^{1}R(0,x_{ps})f_{0}(z,x_{0})\end{equation}
contains the divergence and
\begin{equation}
    \phi_{f}^{R}(x_{0})=\int_{0}^{1}g(z,x_{0})dz\end{equation}
is a regular integral in $z$ and $x_{0}$. To find the order of
divergence of the integrand, we expand the argument of the square
root of $f(z,x_{0})$ to second order in $z$ and get the function
$f_{0}(z,x_{0})$:
\begin{equation}
    f_{0}(z,x_{0}) = \frac{1}{\sqrt{\alpha{z}+\beta{z^{2}}}}, \end{equation}
where:
\begin{eqnarray}
  &&\alpha=sgn(A_{0})\frac{(1-y_{0})}{A_{0}^{\prime}C_{0}}[A_{0}C_{0}^{\prime}-A_{0}^{\prime}C_{0}+J(A_{0}^{\prime}D_{0}-A_{0}D_{0}^{\prime})], \\
  &&\beta=sgn(A_{0})            \frac{(1-y_{0})^{2}}{2C_{0}^{2}A_{0}^{\prime3}}\{2C_{0}C_{0}^{\prime}A_{0}^{\prime2}+(C_{0}C_{0}^{\prime\prime}-2C_{0}^{\prime2})y_{0}A_{0}^{\prime}-C_{0}C_{0}^{\prime}y_{0}A^{\prime\prime} \nonumber \\
 &&\hspace{0.8cm}+J[  A_{0}C_{0}(A_{0}^{\prime\prime}D_{0}^{\prime}-A_{0}^{\prime}D_{0}^{\prime\prime})+2A_{0}^{\prime}C_{0}^{\prime}(A_{0}D_{0}^{\prime}-A_{0}^{\prime}D_{0})]\}.
\end{eqnarray}

The function $g(z,x_{0})$ is the difference between the original
integrand and the divergent integrand
\begin{equation}
  g(z,x_{0})=R(z,x_{0})f(z,x_{0})-R(0,x_{ps})f_{0}(z,x_{0}).
\end{equation}
When $x_{0}$ becomes equal to $x_{ps}$ the equation of photon
sphere holds. Then, the coefficient $\alpha$ vanishes and the
leading term of the divergence in $f_{0}$ is $z^{-1}$. Therefore
the integral diverges logarithmically. The coefficient $\beta$
takes the form
\begin{eqnarray}
\beta_{ps}=sgn(A_{ps})\frac{(1-A_{ps})^{2}}{2C_{ps}{A_{ps}^{\prime}}^{2}}[A_{ps}C_{ps}^{\prime\prime}-A_{ps}^{\prime\prime}C_{ps}+J(A_{ps}^{\prime\prime}D_{ps}-A_{ps}D_{ps}^{\prime\prime})].
\end{eqnarray} Bozza \cite{Bozza1} obtained the analytical
expression of the deflection angle close to the divergence
expanding the two parts of the original integral (\ref{Integral})
around $x_{0}=x_{ps}$ and approximating the leading terms. The
result is
%------------------------------------------------------------------------------------------------------------------------
\begin{figure}
\includegraphics[width=0.49\textwidth]{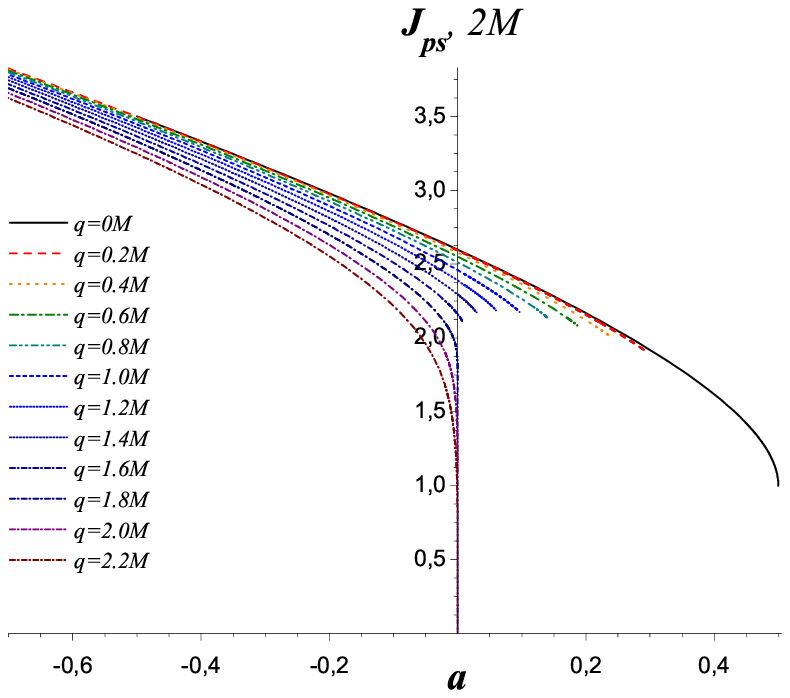}
\includegraphics[width=0.49\textwidth]{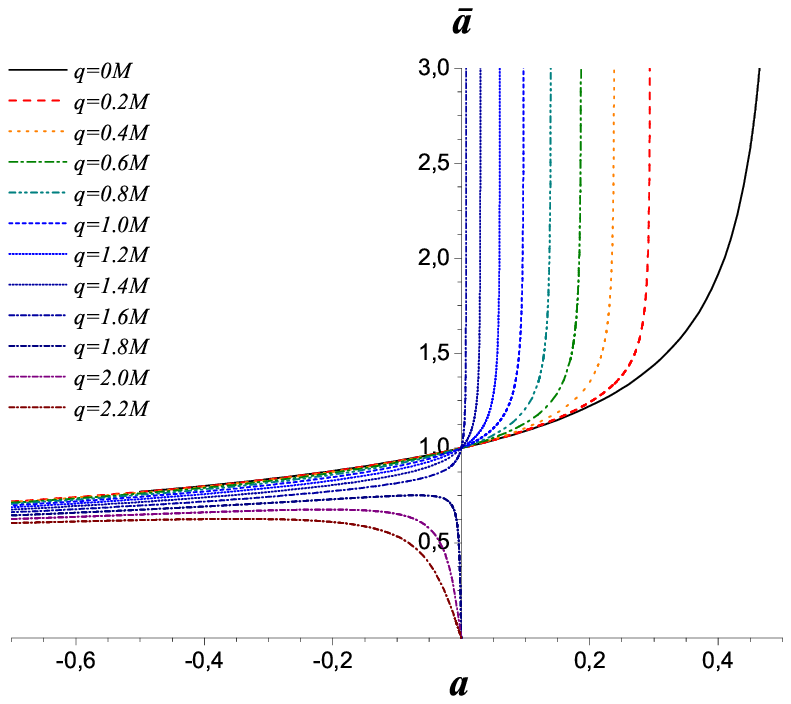}\\
\vspace{0.25cm}
\includegraphics[width=0.49\textwidth]{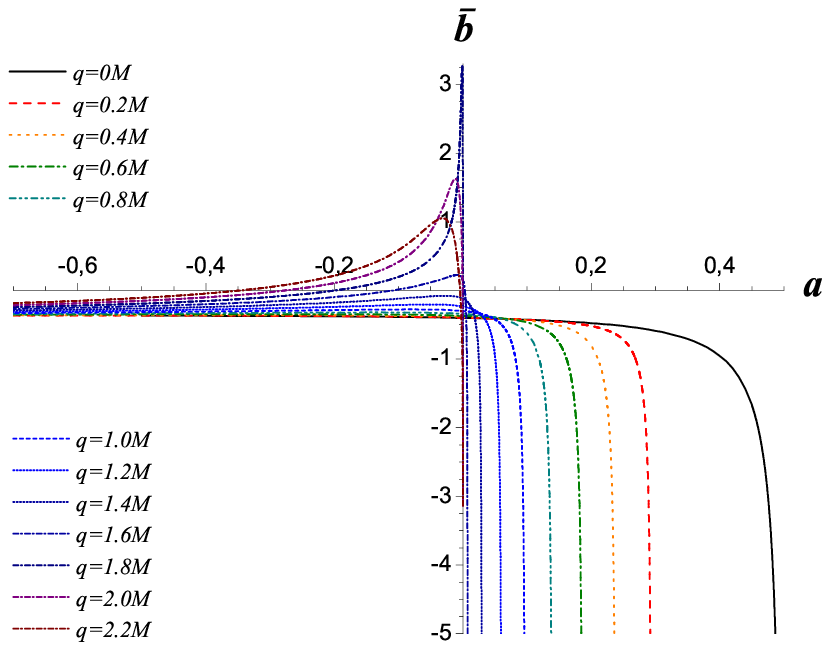}
\caption{\small Strong deflection limit coefficients as a function of the lens angular momentum for scalar charge $q/M=0$ (solid line) to charge $q/M=2.2$ (short, dashed, dotted line) by scalar charge $q/M=0.2$.} \label{u_a_b}
\end{figure}
%------------------------------------------------------------------------------------------------------------------------
\begin{eqnarray}
  \alpha(\theta)=-\bar{a}\ln\left( \frac{\theta{D}_{OL}}{J_{ps}}-1 \right) +\bar{b} +O(J-J_{ps}), \label{DefAng}
\end{eqnarray}
where the coefficients $\bar{a}$, $\bar{b}$ and $J_{ps}$ depend on
the metric function evaluated numerically at $x_{ps}$. $\theta=J/D_{OL}$ is the angular separation between the lens and the image.
The strong deflection limit coefficients of the expansion (\ref{DefAng}) are
\begin{eqnarray}
 &&J(x_{ps})=\frac{a\left(1-\frac{1}{\gamma{x}_{ps}}\right)^{\gamma}-a+\sqrt{x_{ps}^{2}+a^{2}-\frac{x_{ps}}{\gamma}}}{\left(1-\frac{1}{\gamma{x}_{ps}}\right)^{\gamma}},\\
 &&\bar{a}=\frac{R(0,x_{ps})}{2\sqrt{\beta_{ps}}},\\
 &&\bar{b}=-\pi+b_{R}+\bar{a}\ln \left\{\frac{2\beta_{ps}}{y_{ps}}\right\},
\end{eqnarray}
where $b_{R}$ is the regular integral $\phi_{f}^{R}(x_{0})$
evaluated at the point $x_{ps}$ as follows
\begin{equation}
  b_{R}=\int_{0}^{1}g(z,x_{ps})dz.\end{equation}
For the same metrics the coefficient $b_{R}$ cannot be obtained
analytically. In such cases it must be evaluated numerically.

Fig. \ref{u_a_b} shows the strong deflection limit coefficients as
functions of $a$. The minimum impact parameter has a similar
behavior for different scalar charges and decreases with
the increase of $|a|$. $\bar{a}$ grows, while $\bar{b}$ decreases
for $1/2\leq\gamma<1$ and all values of angular momentum. In the
cases $0<\gamma<1/2$ the coefficients $\bar{a}$ and $\bar{b}$
grow with the increase of $-\infty<a<0$ as in the vicinity
of the upper limit the angular momentum $\bar{a}$ vanishes and
$\bar{b}$ becomes $-\pi$. Both coefficients diverge in the
vicinity of the upper limit of the Kerr black hole and WNS angular
momentum, where the strong deflection limit approximation fails. The
divergence of the coefficients of the expansion means that the
bending angle in the strong deflection limit (\ref{DefAng}) no
longer represents a reliable description in the regime of high $a$.

\subsection{Observables in the strong deflection limit}

As already was marked the lens equation (\ref{LensEq1}) can reduces to lens equation describing large deflection angle of the light ray when the measured from the observer image and source angles are small with respect to the optical axis. Taking the general lens geometry in consideration, and using the strong deflection limit lens equation
\begin{eqnarray}
  &&\eta=\frac{D_{OL}+D_{LS}}{D_{LS}}\theta-\alpha(\theta) \hspace{0.2cm} mod \hspace{0.2cm} 2\pi,  \\
  &&\alpha(\theta) = -\bar{a}\ln\left( \frac{\theta{D}_{OL}}{J_{ps}}-1 \right) +\bar{b}, \label{LensEq2}
\end{eqnarray}
one can show \cite{BozQuazi} that the angular separation between the
lens and the n-th relativistic image is
\begin{eqnarray}
  \theta_{n}^{\pm}\simeq\pm\theta_{n}^{0}\left(1-\frac{J_{ps}e_{n}^{\pm}(D_{OL}+D_{LS})}{\bar{a}D_{OL}D_{LS}}\right),
\end{eqnarray}
where
\begin{eqnarray}
  \pm\theta_{n}^{0} = \frac{J_{ps}}{D_{OL}}(1+e_{n}^{\pm}), \hspace{0.5cm} e_{n}^{\pm}=e^{\frac{\bar{b}\pm|\eta|-2\pi{n}}{\bar{a}}}.
\end{eqnarray}

Here $\eta$ is the angular separation between the source and the optical axis, as seen from the lens. According to the past oriented light ray which starts from the observer and finishes at the source the resulting relativistic images $\theta_{n}^{+}$ stand on the eastern side of the black hole for direct photons ($a>0$) and are described by positive $\eta$. Retrograde photons ($a<0$) have relativistic images $\theta_{n}^{-}$ on the western side of the black hole and are described by negative values of $\eta$. $n$ is the number of loops done by the photon around the lens object. For each $n$, we have an image on each side of the lens, according to the chosen sign. We have expressed the position of the relativistic images in terms of the coefficients $\bar{a}$, $\bar{b}$ and $J_{ps}$. If we manage to determine these coefficients from the observation of the relativistic images, we can obtain information about the parameters of the lens object stored in them.
%----------------------------------------------------------------------------------------------------------------------
\begin{figure}
\includegraphics[width=0.49\textwidth]{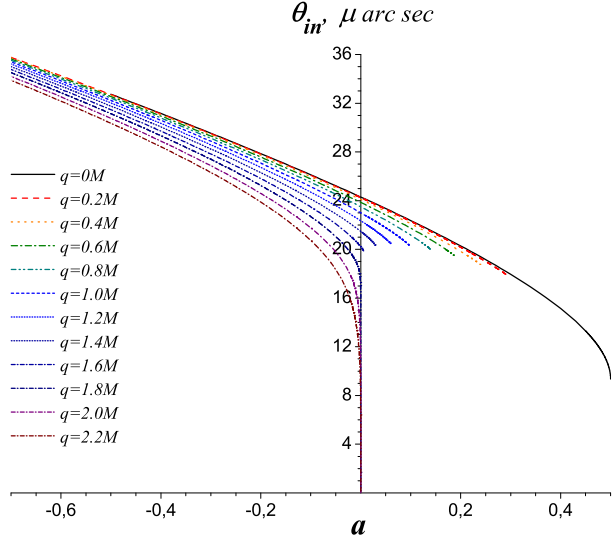}
\caption{\small The innermost relativistic image $\theta_{in}=\theta_{n}{|}_{n\rightarrow\infty}\equiv\theta_{\infty}$ for scalar charge $q/M=0$ (solid line) to charge $q/M=2.2$ (short, dashed, dotted line) by scalar charge $q/M=0.2$.} \label{Observables1}
\end{figure}
\begin{figure}
\includegraphics[width=0.50\textwidth]{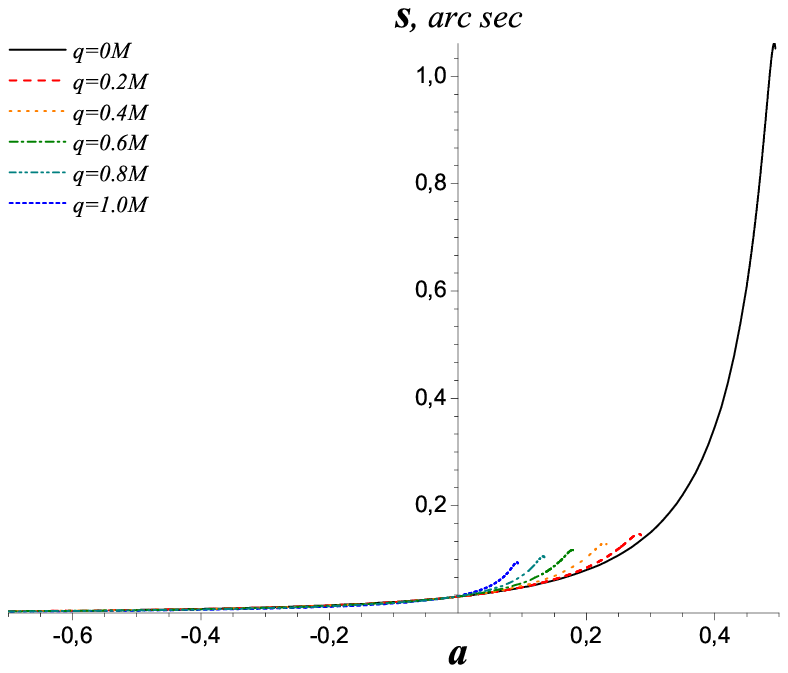}
\includegraphics[width=0.49\textwidth]{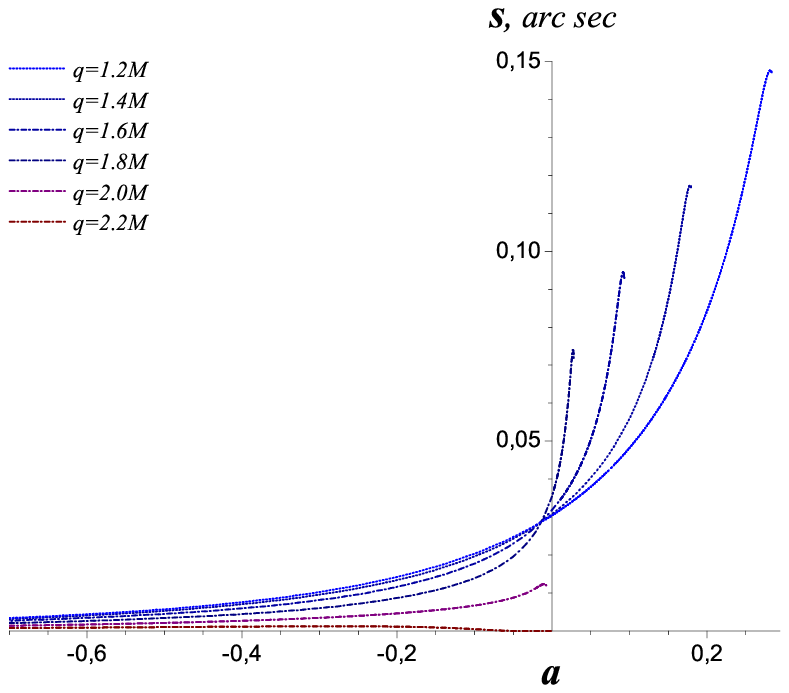}
%\resizebox{5.5cm}{4.5cm}{\includegraphics{file/s.eps}}\\
\caption{\small The separation between the outermost relativistic image $\theta_{1}$ and all the remaining ones packet together at $\theta_{\infty}$ for scalar charge $q/M=0$ (solid line) to scalar charge $q/M=1.0M$ (short, dashed line) (on the left) and for scalar charge $q/M=1.2M$ to scalar charge $q/M=2.2$ (short, dashed, dotted line) (on the right) all by scalar charge $q/M=0.2$.} \label{Observables2}
\end{figure}
%----------------------------------------------------------------------------------------------------------------------

\newpage

To obtain the coefficients $\bar{a}$, $\bar{b}$, we need to separate at least the outermost image from all the others.
If we consider the simplest situation where only the outermost image $\theta_{1}$ is resolved as a single image,
while all the remaining ones are packed together at $\theta_{\infty}$, our observables will thus be \cite{Bozza1}
\begin{eqnarray}
  s=\theta_{1}-\theta_{\infty}=\theta_{\infty}e^{\bar{b}/\bar{a}-2\pi/\bar{a}}.
\end{eqnarray}
which represent the separation between the
first image and the others. The innermost relativistic images and the separability are computed numerically for different values of the scalar charge $q/M$ and the lens angular momentum and are plotted in the Fig. \ref{Observables1} and Fig. \ref{Observables2} for the massive compact object in the center of our Galaxy \cite{Eis}.

The gap between the outermost relativistic image and the remaining ones packed together
at inner image position increases with the increase of
$|a|$ for all values of the scalar charge. It is obvious that in the
vicinity at $a=0$, where the SDL approximation works sufficiently,
the WNS separability $s$ exceeds the Kerr black hole
separability. While for bigger positive angular momentum the
dependence is the same, for negative angular momentum, a value
$\acute{a}$ exist after which the image separability is lower in
comparison to the Kerr black hole separability. The graphics are
truncated in the vicinity of the separable point where the SDL
approximation fails.

Due to the non-existence of separation of the variables in the rotating singularity space-time we have not examined the magnification of the strongly demagnified relativistic images.

\section{Discussion and Conclusion}

In this work we discussed the features of light propagation
in the spacetime of a stationary,
axially-symmetric black hole and naked singularities. Modelling the
massive compact object in the center of the galaxy
as a rotating generalization of the Janis--Newman--Winicour
naked singularity, we estimate the numerical values of the
deflection angle of the light ray for the Kerr black hole, WNS and
SNS. Provoking from the weak deflection limit analysis, basing on the degeneracy of a rotating singularity lens and a displaced  Janis--Newman--Winicour lens we derived analytical descriptions for the two weak field images, the critical curves, the caustic points, the signed magnifications and the total magnification as well as the magnification centroid up to post-Newtonian order. Moreover, in our striving to study the gravitational lensing in the strong deflection limit, we evaluated the strong deflection limit coefficients of the deflection angle for the WNS, and the lensing observables ensuing from them. When comparing the results to the corresponding quantities for a rotating naked singularity and Kerr black hole, we find that there are rotation and scalar field effects present in the behavior of the
bending angle of the light ray, as well as the lensing observables.

The analytical results show that there are a static post-Newtonian corrections to the two weak field image positions, to the signed magnification and to their sum as well as to the critical curves, which are function of the scalar charge. The shift of the critical curves as a function of the lens angular momentum is found, and it is shown that they decrease slightingly for the weakly naked and vastly for the strongly naked singularities with the increase of the scalar charge. The point-like caustics drift away from the optical axis and do not depend on the scalar charge. In the static case of Janis--Newman--Winicour lensing the scalar charge leads to increases of the signed magnifications for arbitrary source position. However, when the source is equatorial (\textit{i.e.} $\beta_{2}=0$), with the increase of the scalar charge the total magnification $\mu_{tot}$ decreases for a source on the right hand side of the optical axis (\textit{i.e.} $\beta_{1}>0$) and increases when the source is on the left hand side (\textit{i.e.} $\beta_{1}<0$). The non-equatorial source position leads only to an increase of the total magnification for fixed $q/M$.

In the case of rotating naked singilarity lensing, the point-like caustic $\beta_{1}^{cau}$ drift away from the optical axis and some corrections with respect to Janis--Newman--Winicour lensing have arisen. When the source is equatorial, for every values of $\beta_{1}$, the positive parity $\mu_{+}$ and negative parity $\mu_{-}$ magnifications respectively increases and decreases with the increase of $q/M$ with respect to signed magnifications in the static case. The non-equatorial source position leads to an increase of $\mu_{+}$ for $\beta_{1}>0$ and to decrease of it for $\beta_{1}<0$ with respect to the static case. $\mu_{-}$ has a opposite behavior with respect to $\mu_{+}$. In this cases the scalar charge increases the signed magnifications for all values of lens angular momentum $a$ and source positions $(\beta_{1}$, $\beta_{2})$. The total magnification remind the behavior of positive parity magnification $\mu_{+}$, with the difference that for a fixed $a$ with the increase of the scalar charge $\mu_{tot}$ increases for $\beta_{1}>0$ and decreases for $\beta_{1}<0$. Looking from the observer position the magnification centroid $\Theta^{\rm Cent}$ seem like a bell-like curve for a non-equatorial source and like a straight line for a equatorial source. The lens angular momentum $a$ leads to an increase of the centroid for $\beta_{1}>0$ and to decrease of it for $\beta_{1}<0$.

Let us to resume the strong deflection limit results. Because of the existence of the photon sphere in the
case of the Kerr black hole and rotating WNS, gravitational lensing
gives rise to a sequence of an infinite number of highly demagnified
relativistic images distributed respectively on the each side of the
optical axis. In the case of
Janis--Newman--Winicour SNS when $a=0$ and $(q/M)^{2}\geq3$ the
gravitational lensing does not give relativistic images because of
the non-existence of a photon sphere. In the rotational case of WNS the position
of the relativistic images decreases with the increase of
$|a|$ for all values of the scalar charge and all of
the images are closer to the center of the lens in comparison to
the Kerr images. For all values of $q/M$ the relativistic images are
closer to the optical axis for positive $a$ in the case of direct
photons and further from the optical axis for negative $a$ when we
consider opposite photons. For the static case when
Janis--Newman--Winicour WNS are realized the relativistic images are
not shifted. Therefore as seen from the side of the observer the
relativistic images are shifted towards the western side, if north
is the direction of the spin. The gap between the
outermost relativistic image and the remaining ones packed together
at inner image position increases with the increase of
$|a|$ for all values of the scalar charge. In the
vicinity at $a=0$, where the SDL approximation works sufficiently,
the WNS separability $s$ exceeds the Kerr black hole
separability. While for bigger positive angular momentum the
dependence is the same, for negative angular momentum, a value
$\acute{a}$ exist after which the image separability is lower in
comparison to the Kerr black hole separability.

According to the results, the gravitational lensing
in the weak and the strong deflection limit would allow us to distinguish the
Kerr black hole from rotating weakly naked singularities. Hence,
detecting the weak field or/and the relativistic images, which might be possible in the
near future, we will be able to determine the nature of the lensing
galactic massive dark object.

\begin{acknowledgments}
We wish to express our gratitude to K. S. Virbhadra and V. Bozza for helpful correspondence, useful discussions, and nice suggestions. This work was partially supported by the Bulgarian National Science Fund under Grants MUF 04/05 (MU 408), VUF-201/06 and VUF-205/06, and the Sofia University Research Fund No. 111.
\end{acknowledgments}

\end{document}